# SHARP—A spectrograph proposal to fully exploit ELT capabilities and look beyond JWST ☆,☆☆

P. Saracco [a],*, P. Conconi [a], H. Mahmoodzadeh [a,b], C. Arcidiacono [c], I. Di Antonio [d], E. Portaluri [d], P. Franzetti [e], A. Gargiulo [e], I. Arosio [a], L. Barbalini [a], G. Lops [a], E. Molinari [a], J.M. Alcalá [f], S. Bisogni [e], R. Bonito [g], E. Bortolas [c], M. Cantiello [d], A. Caratti o Garatti [f], E. Cascone [f], V. Cianniello [f], E.M. Corsini [h], F. Damiani [g], F. D'Ammando [i], F. D'Alessio [j], E. Dalla Bontà [h], M. Dall'Ora [f], V. De Caprio [f], G. De Lucia [k], B. Di Francesco [d], G. Di Rico [d], V. D'Orazi [l], C. Eredia [f], A.R. Gallazzi [m], M.G. Guarcello [g], L. Izzo [f], F. La Barbera [f], M. Lippi [m], M. Longhetti [a], A. Longobardo [n], C. Mancini [e], M. Mirabile [d], E. Piconcelli [j], A. Pizzella [h], L. Podio [m], L. Prisinzano [g], C. Tortora [f], G. Vietri [e], F. Vitali [j], H.-F. Wang [h], S. Zibetti [m]

[a] *INAF - Osservatorio Astronomico di Brera, via Brera 28, 20121 Milano, Italy*
[b] *Politecnico di Milano, Piazza Leonardo da Vinci 32, Milano, 20133, Italy*
[c] *INAF - Osservatorio Astronomico di Padova, Vicolo Osservatorio 5, Padova, 35122, Italy*
[d] *INAF - Osservatorio Astronomico d'Abruzzo, Via Mentore Maggini snc, Teramo, 64100, Italy*
[e] *INAF - IASF, Via Alfonso Corti 12, Milano, 20133, Italy*
[f] *INAF - Osservatorio Astronomico di Capodimonte, Salita Moiariello 16, Napoli, 80131, Italy*
[g] *INAF - Osservatorio Astronomico di Palermo, Piazza del Parlamento 1, Palermo, 90134, Italy*
[h] *Universita' degli Studi di Padova, Via F. Marzolo 8, Padova, 35131, Italy*
[i] *INAF - Istituto di Radioastronomia, via Gobetti 101, Bologna, 40129, Italy*
[j] *INAF - Osservatorio Astronomico di Roma, Via Frascati 33, Monte Porzio Catone (RM), 00078, Italy*
[k] *INAF - Osservatorio Astronomico di Trieste, Via G.B. Tiepolo, 11 I-Trieste, 34143, Italy*
[l] *Universita' degli Studi di Roma Tor Vergata, Via Cracovia 50, Roma, 00133, Italy*
[m] *INAF - Osservatorio Astrofisico di Arcetri, Largo E. Fermi 5, Firenze, 50125, Italy*
[n] *INAF - IAPS, Via del Fosso del Cavaliere 100, Roma, 00133, Italy*

## ARTICLE INFO

*Keywords:*
ELT
Near-infrared
Multi-object spectrograph
Integral field spectroscopy
Galaxies
Evolution
Formation

## ABSTRACT

The Extremely Large Telescopes (ELTs), with their large apertures and cutting-edge Multi-Conjugate Adaptive Optics (MCAO) systems, promise to deliver data that is both sharper and deeper than even the James Webb Space Telescope (JWST) across large fields. SHARP is a concept study for a near-IR (0.95–2.45 μm) spectrograph specifically designed to fully exploit the collecting area and angular resolution capabilities of the upcoming ESO's ELT. The instrument concept is driven by the goal of tackling the most important questions in astrophysics and cosmology, from exploring primordial galaxies to studying the formation of young stellar object and planetary systems in the nearby dust-enshrouded regions, bridging the gap between the local and the distant Universe. This requires versatility to accommodate diverse observational needs. SHARP is composed of two main units: NEXUS, a Multi-Object Spectrograph (MOS) optimized for detecting the faintest sources, and VESPER, a multi-object Integral Field Unit (multi-IFU) designed for brighter ones. This article provides an overview of the scientific design drivers, the solutions developed to meet them, and the resulting optical design that achieves the required performance.

☆ This article is part of a Special issue entitled: 'SHARP Science Book' published in New Astronomy.
☆☆ SHARP website http://sharp.brera.inaf.it.

* Corresponding author.
*E-mail address:* paolo.saracco@inaf.it (P. Saracco).

## 1. Introduction

The upcoming generation of ground-based Extremely Large Telescopes (ELTs) will peer into the Universe with unparalleled clarity delivering sharper and deeper data than the James Webb Space Telescope (JWST, Gardner et al., 2023). These capabilities are due to their large apertures providing a large photon collecting area and to advanced multi-conjugate adaptive optics (MCAO) systems designed to uniformly correct large fields for atmospheric turbulence. The ELTs will achieve data significantly sharper than JWST, up to four times in the case of the Giant Magellan Telescope (GMT, Johns, 2006; Burgett et al., 2024) (∼25 m diameter), and six times sharper with the ESO's Extremely Large Telescope (ELT, Gilmozzi and Spyromilio, 2007; Vernet et al., 2024).

In particular, ESO's ELT is positioned to set new standards in astronomical observations, thanks to its 39 m diameter primary mirror, which will be the largest ever constructed, and its advanced MCAO system MORFEO (Ciliegi et al., 2021, 2024). MCAO systems are designed to provide uniform correction for atmospheric turbulence across a wide area. For instance, MORFEO corrects an area with a diameter of 160 arcsec, an AO system that allows supported instruments to explore the Universe with unmatched depth and clarity. MORFEO will support MICADO (Davies et al., 2021, 2018), the ELT's first-light near-IR camera and long-slit spectrograph, and it is also configured to serve a second instrument, HARMONI (Thatte et al., 2024), a single-field IFU. A spectrograph that fully exploits the capabilities of an MCAO system like MORFEO at the ELT, namely, the large AO corrected area and the faint achievable flux limit, is expected to surpass NIRSpec (Jakobsen et al., 2022) at JWST in terms of spectral and angular resolution, allowing us to continue and expand on JWST's discoveries once its mission is completed.

SHARP[1] is a concept study for a multi-mode near-IR spectrograph specifically designed to leverage the capabilities of MCAO-assisted observations at the ELT. It is conceived to make the most of both the large aperture of the ELT and the characteristics of an MCAO system as MORFEO, namely the high angular resolution over such a wide field. This would provide an effective spectroscopic follow-up of the deep and high angular resolution fields expected from MICADO observations. Developed as a candidate for the future call from ESO for new instrumentation and observing facilities,[2] SHARP is engineered to exploit the full potential of MCAO-assisted observations, offering an angular resolution and a range of spectral resolutions that go well beyond those provided by NIRSpec at the JWST.

This article provides an overview of the scientific drivers behind SHARP, the key requirements that arise from them, and the resulting optical design that can achieve the required performance. In Section 2 we define the scientific context in which SHARP fits and provide an overview of the main scientific drivers underlying the SHARP concept. Section 3 summarizes the requirements arising from the scientific context. Section 4 describes the overall SHARP architecture, details the two main units, NEXUS and VESPER, the multi-object spectrograph and the multi-integral field unit, respectively and the relevant subsystems. Finally, Section 5 shows a representative example of SHARP observation and its returns with reference to some of the scientific drivers discussed in Section 2.

Throughout this paper, we adopt $H_0 = 70$ km/s/Mpc, $\Omega_m = 0.3$ and $\Omega_\Lambda = 0.7$.

[1] http://sharp.brera.inaf.it

[2] Expanding Horizons, https://next.eso.org/

## 2. Scientific context and drivers

In this section, we will present the main scientific drivers behind the development of SHARP, without detailing the individual science cases. These specific cases will be thoroughly addressed in the various articles that constitute this SHARP Science Book special issue[3].

One of the main scientific driver originally behind the development of a spectrograph like SHARP is understanding how baryons assembled to form the first stars, galaxies, and cosmic structures, and how these systems subsequently evolved over cosmic time. These fundamental themes and questions are central to current astrophysics, as they essentially map the path of cosmic evolution to understanding the Universe at the various scales as we observe it today.

### 2.1. Formation and evolution of galaxies

A deep understanding of galaxy life-cycle requires understanding the star formation processes, the mechanisms at play, whether they are affected by the environment and, if so, what are the physical processes that intervene. Clarity is needed on the mechanisms that regulate the shutdown of star formation, often capable of abruptly stopping star formation rates of several hundred solar masses per year (e.g., D'Eugenio et al., 2020; Pérez-González et al., 2025). Furthermore, there is no single universally accepted cause for quenching. Instead, the current consensus is that quenching may result from different possible physical mechanisms broadly divided into two main categories (see Peng et al., 2010, and references therein): internal mechanisms (AGN and star formation feedback, morphological and halo quenching) and external mechanisms (ram-pressure stripping, strangulation/starvation). Analogously, it is not clear what keeps a galaxy in a quiescent state thereafter despite the presence, in many cases, of large amounts of gas or even massive gas inflow (e.g., Bevacqua et al., 2026). In this regard, the connection between galaxy life-cycle and the intergalactic (IGM) and circumgalactic (CGM) medium is still rather unclear (e.g., Looser et al., 2024).

#### 2.1.1. Massive galaxies

In the current model of galaxy formation dark matter (DM) accumulated in gravitationally bound halos forming the seeds of the first galaxies. These would grow over time through baryon accretion, star formation, and merging with other DM halos (e.g., Springel et al., 2005). In this hierarchical paradigm, massive galaxies form through the merger of smaller primordial galaxies, completing their assembly at later times and hence becoming common in the local Universe, increasingly rare at higher redshifts, and disappearing in the early cosmic epochs (e.g., De Lucia et al., 2006; de Lucia, 2013). Therefore, the search for massive galaxies at high redshifts has been a focus of attention for decades, as it represents a powerful test of the hierarchical model (e.g., Thompson et al., 1999; Daddi et al., 2000; McCarthy et al., 2001; Cimatti et al., 2002; Saracco et al., 2003). The main physical processes and mechanisms involved in this scenario were thought to be fairly well understood. However, deep observations accumulated over the past decade have revealed numerous massive galaxies at high redshift, challenging our understanding of galaxy formation physics. In practice, they appear to be too massive, too numerous, and too early in cosmic timescales compared to theoretical expectations (e.g., Boylan-Kolchin, 2023; Chworowsky et al., 2024; Carnall et al., 2024; Glazebrook et al., 2024; Xiao et al., 2024).

[3] https://www.sciencedirect.com/special-issue/10W3B6JJBXH

*Star formation* - JWST has detected massive quiescent galaxies up to $z \sim 5$, when the Universe was younger than 1.2 Gyr. These include both galaxies whose stellar population is extremely old, that is, formed a few million years after the Big Bang, and galaxies whose stars are much younger, having formed a few million years before observation (e.g., Glazebrook et al., 2017; Tanaka et al., 2019; Saracco et al., 2020; D'Eugenio et al., 2020; Antwi-Danso et al., 2025; Carnall et al., 2024; Glazebrook et al., 2024). Given these high stellar masses (log(M*/M⊙)∼11) and the few million years available to form them, the resulting star formation rates (SFR) must be higher than hundreds solar masses per year. These values are particularly challenging under the conditions of the early Universe: the efficiency of star formation is expected to be reduced by several factors, including the lower cooling efficiency of pristine hydrogen, the higher UV radiation density and pressure (due to the Universe's greater density), and the fact that Dark Matter (DM) halos of sufficient hosting mass were not yet formed (e.g., Boylan-Kolchin, 2023; Glazebrook et al., 2024). However, it is worth mentioning that Krishnan and Abazajian (2026) show that this apparent tension could be driven by the strong coupling of systematic uncertainties in stellar mass estimates from the SED fit with the exponentially falling dark matter halo mass function, according to which it is much more likely that we are overestimating the stellar mass of a low-mass galaxy at early Universe than vice versa. The authors show that if these SED systematics and asymmetric scatter are properly accounted for, the required star formation efficiencies drop to physically plausible levels, largely reducing the tension.

*Metallicity and enrichment* - Massive galaxies, whether they host old or young stellar populations, are characterized by solar metallicity or higher, e.g., [Fe/H]∼0.02 (Glazebrook et al., 2024) and [Z/H]>0.15 (Saracco et al., 2020). Such high metallicity values further challenge theories of chemical enrichment, since such levels require times (> 1 Gyr) apparently inconsistent with those of the star formation times deduced for these galaxies and/or with initial pristine hydrogen. However, the dependence of metallicity and enrichment timescale on the poorly known stellar initial mass function (IMF, see later in this paper) in high redshift galaxies makes this statement uncertain.

*Quenching and quiescence* - Since both massive high-redshift galaxies with young stellar populations and those with old stellar populations appear quiescent, with virtually no ongoing star formation, an extremely efficient shutdown mechanism must have taken place, capable of abruptly halting star formation rates of hundreds of solar masses per year. Quenching mechanisms alternative to AGN outflows are needed, since signs of outflow (asymmetric absorption lines) are seen in very few fast-quenched post-starburst galaxies observed at intermediate and high redshift. Moreover, outflows seem to be not efficient in removing gas, the only way so far known capable of abruptly halting star formation (Lamperti et al., 2021; Concas et al., 2022).

Also, it is reasonable to assume that most of the massive galaxies will remain quiescent, without experiencing further significant episodes of star formation, to prevent them from exceeding the mass (and age) limits of galaxies observed in the local Universe. This assumption holds despite the presence in many of them of a significant quantity of residual gas and/or massive gas inflows (e.g., Belli et al., 2024; Bevacqua et al., 2026). The reason why these galaxies fail to ignite subsequent star formation in such gas-rich environments remains an open question.

*Mass growth and hierarchical assembly* - Indeed, massive, high-redshift ($z \sim$3-4) galaxies that host old (∼1.5–2 Gyr) stellar populations may have accumulated their mass through the merger of smaller, primordial galaxies, since time could be sufficient. Therefore, their presence could be still consistent with the hierarchical paradigm, even if this scenario still faces the challenge of explaining the high star formation rate required to account for the large stellar mass formed in such a short cosmic time even if spread among multiple galaxies. On the contrary, the presence of massive galaxies at high redshift hosting young (∼0.5 Gyr) stellar populations, i.e., formed few million years before the observations, challenges hierarchical paradigm because the stellar mass must have formed in situ (e.g., Puskás et al., 2025), given that there is insufficient time for assembly through the merging of individual subunits (Boylan-Kolchin et al., 2008; Boylan-Kolchin, 2023). Moreover, since these galaxies are in a quiescent state, a mechanism capable of rapidly and abruptly shutdown star formation must have taken place.

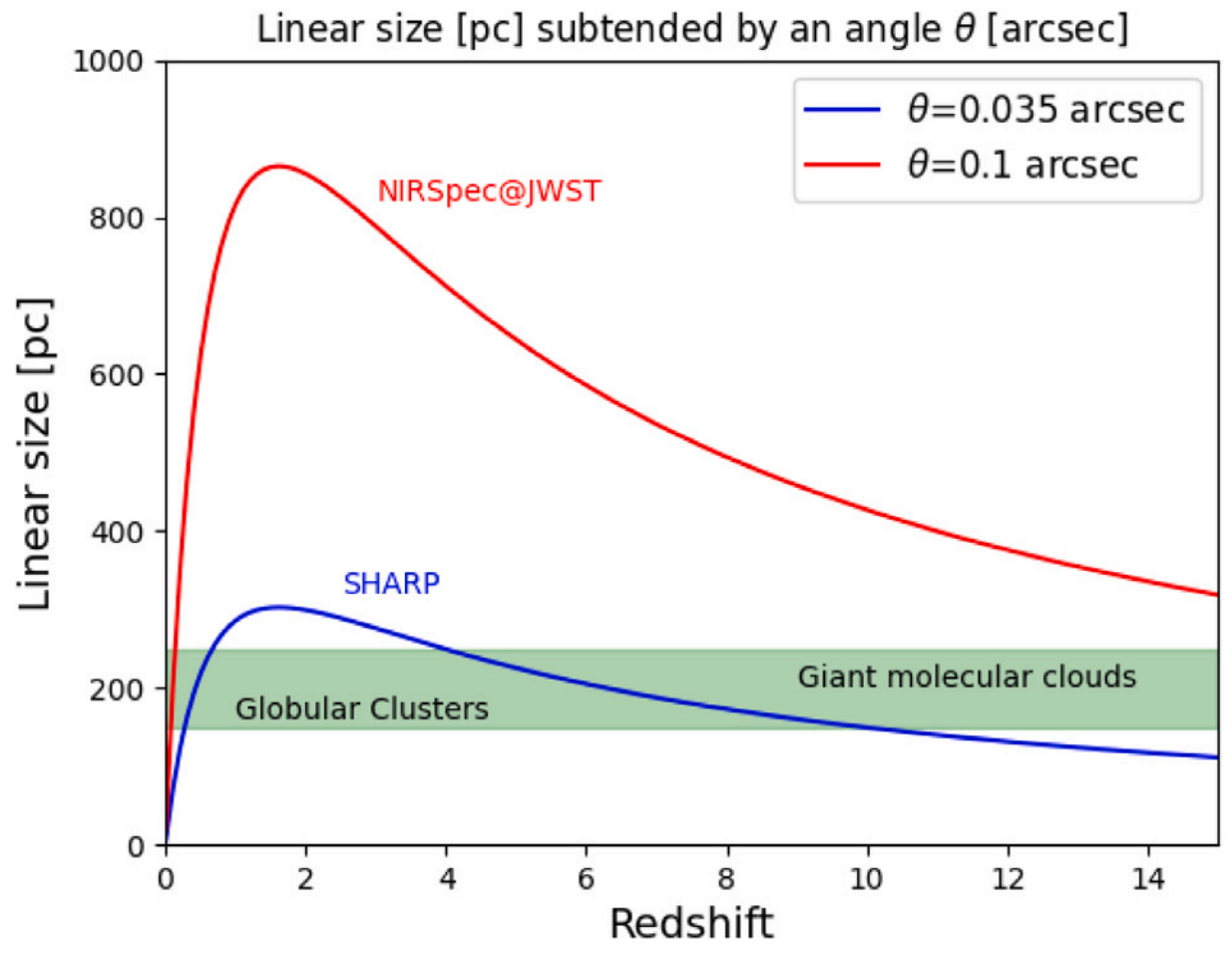

**Fig. 1.** Linear size [pc] in subtended by a pixel of SHARP, $\theta = 0.035''$ (blue band), as a function of redshift. The green horizontal stripe represents the typical size of giant molecular gas clouds, 150–250 pc. For comparison, the red line is the linear size subtended by the pixel size (0.1"/pix) of NIRSpec at the JWST.

*Observational needs and challenges* - High-redshift massive galaxies, regardless of whether they host old or young stellar populations, present significant challenges to hierarchical models due to the necessity of:

- inexplicably high/efficient star formation rates
- unexpected fast and efficient chemical enrichment
- extremely rapid quenching mechanisms.

To tackle these issues, it is essential to study the processes of star formation, chemical enrichment, and quenching in a homogeneous way up to high redshift. This requires capturing the physical properties at the scale of star-forming regions within high-redshift galaxies to relate the increased SF efficiency, enrichment, and rapid quenching to the properties of the environment in which star-forming regions are located. As discussed in Section 3.1 massive baryons assembly in galaxies takes place within Giant Molecular gas Clouds (GMCs). This clouds have typical size of ∼200 pc requiring an angular resolution of about 0.03" to be resolved over the whole cosmic time (see Fig. 1). It worth to mention that such physical scales are sampled by NIRSpec at JWST at redshift $z < 0.1$.

These spectroscopic measurements, which require near-IR coverage (see Figs. 2 and 3), are currently beyond the capabilities of JWST due to its limited angular resolution (see Fig. 1). The capabilities of next-generation ELT spectrographs are affected, in some cases, by limitations in angular resolution and limited wavelength coverage, in other cases by a limited field of view or observing mode that prevents them from observing more than one source simultaneously.

Therefore, from an observational standpoint, high sensitivity, high angular resolution and multiplexing capabilities are mandatory. This allows for the homogeneous study of the physical conditions under which star formation, enrichment, and quenching occur at different redshifts within galaxies.

The articles by Gargiulo et al. (2026), Mancini et al. (2026) Longhetti et al. and Crescenzo et al. in this special issue address in detail some of the issues described above.

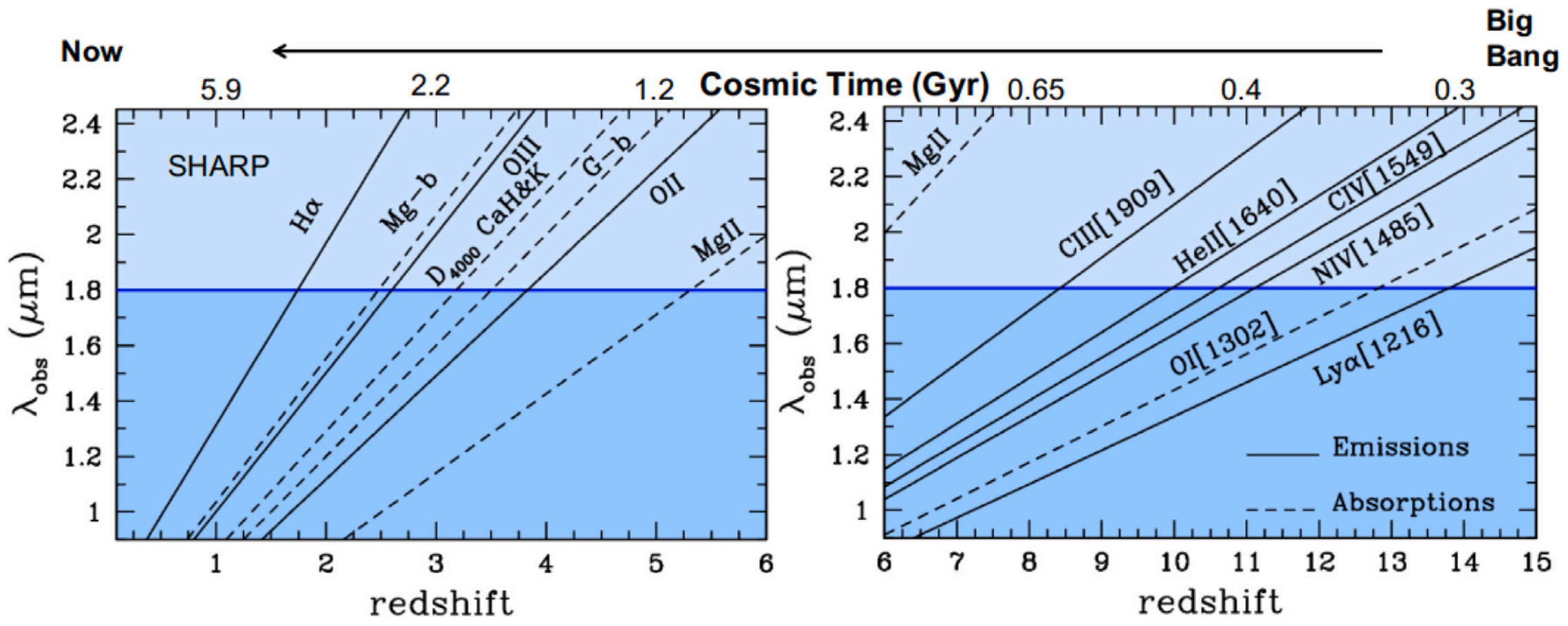

**Fig. 2.** Observed wavelength of the main atomic emission (solid lines) and absorption lines (dotted lines) as a function of redshift. The blue horizontal line marks the limiting wavelength at 1.8 μm of some of the next generation spectrograph as MOSAIC and ANDES at the ELT and MOONS at the VLT. In the left-hand panel, the atomic lines in the visible rest-frame (0.35–0.65 μm), tracing stellar population properties and kinematics of galaxies, are shown. In the right-hand panel, the atomic lines in the UV rest-frame wavelength 0.12–0.30 μm tracing intergalactic medium and early Universe properties.

#### *2.1.2. Scaling relations*

Galaxies follow key empirical scaling relations that link their structural and kinematic properties (such as effective radius $R_e$, stellar velocity dispersion $\sigma_e$, and mass) to those of their stellar populations (such as age, metallicity $Z$, and IMF slope). The nature of these relations, whether they are the result of later evolution or direct formation imprints, is yet unclear. Hence, their significance lies in their ability to strongly constrain and discriminate between competing evolutionary and formation scenarios, as not all theoretical processes can reproduce the observed relations. Therefore, tracing both stellar population and structural properties back through cosmic time is essential to leverage these relations as reliable constraints.

Among the main relationships are the Fundamental Plane (FP) (Djorgovski and Davis, 1987; Dressler et al., 1987), which connects $\sigma_e$ and $R_e$ to the SB, the age-mass and metallicity-mass relations (e.g., Gallazzi et al., 2005) which connect the age and the metallicity Z to the stellar mass of the galaxy. When early-type galaxies are plotted on the three-dimensional space [$\sigma_e$, $R_e$, SB] they distribute in such a way to define a plane (e.g., Saglia et al., 2016; Beifiori et al., 2017; Jørgensen and Chiboucas, 2013; Holden et al., 2010). The remarkably small intrinsic dispersion of this relation, traditionally interpreted as evidence for a coeval formation epoch for these galaxies, represents a significant challenge to the hierarchical paradigm: the growth through the stochastic process of successive mergers likely produces a higher dispersion compared to the observed one.

Similarly, when the mean age and metallicity of galaxies are plotted as a function of their mass, the average behavior is that the higher the mass, the older and more metal-rich the stellar population, regardless of their environment and at least up to $z \sim 1.5$ (e.g., Jørgensen et al., 2014; Saracco et al., 2023; Thomas et al., 2005, 2010; Choi et al., 2014; McDermid et al., 2015). These age-mass and Z-mass relations presents a conceptual challenge, as metal production requires time, which conflicts with the old age of the stars that requires a rapid star formation for massive high-redshift galaxies. Furthermore, a mass-dependent dispersion of age and Z suggests a more complex picture than the average behavior of the relation (Bevacqua et al., 2024), the nature of which requires a much more detailed measurements of the properties of galaxies.

*Observational needs and challenges* - Addressing the nature of scaling relations requires, on the one hand, integrated measurements for galaxies at different redshifts to test whether scaling relations arise from evolution or formation, and, on the other hand, spatially resolved measurements of properties within galaxies to relate the properties to processes that may have occurred. It remains crucial to understand the processes of star formation and chemical enrichment, and how these fundamental processes depend on other factors such as the galactic environment, gas inflow/outflow, and feedback mechanisms. An important (unknown) parameter that can play a role is the stellar initial mass function, which can significantly influence enrichment and its timescales. For the reasons outlined above, most of these measurements are currently beyond the capabilities of the JWST and the planned ELT spectrographs (see also Section 5).

#### *2.1.3. The stellar initial mass function*

The stellar initial mass function (IMF) defines the distribution of stellar masses at birth within a population. Its knowledge is a cornerstone of astrophysics since it defines galaxy mass scale, govern stellar feedback strength, and shapes chemical enrichment. Traditionally assumed to be universal (matching the solar neighborhood), a growing body of observational evidence now suggests the IMF varies across different environments, both in star-forming and quiescent galaxies (e.g., Hopkins, 2018; Smith, 2020). Understanding the physical drivers behind these variations has become one of the central challenges in modern galaxy formation and evolution studies.

By targeting IMF-sensitive absorption features in nearby early-type galaxies, such as the gravity-dependent Na I doublet at $\lambda\lambda$8183,8195 Å (NaI 8190) and the Wing-Ford band FeH0.99 ($\lambda$~9900 Å), several studies consistently found that the central regions of the most massive systems host a dwarf-enriched (bottom-heavy) IMF (e.g., Conroy and van Dokkum, 2012; La Barbera et al., 2013; Spiniello et al., 2014; La Barbera et al., 2019; Parikh et al., 2018). Studying IMF gradients can help distinguish between in situ star formation and stellar mass assembly via mergers, the latter of which is expected to flatten the gradients. Furthermore, radial variations may indicate that the physics of cloud fragmentation and star formation may depend on local environmental conditions. In central regions with higher density than in the peripheries, for example, the gas might fragment differently, leading to the formation of a greater number of stars with a given mass. Studies of this kind become crucial when conducted in high-redshift galaxies. This is because such galaxies are expected to have experienced significantly fewer merging events (if any) compared to their local counterparts. In this case, the observed IMF gradients provide a direct indication of the dependence of star formation on local environmental conditions within the galaxy.

The paper by La Barbera et al. (2026) in this special issue addresses the problem of IMF variation both radially within galaxies and between galaxies (see also the paper by Gargiulo et al., 2026, on stellar population gradients).

*Observational needs and challenges* - Probing IMF radial gradients at high redshift necessitates sampling scales smaller than 500 pc, considering that the most compact massive galaxies observed at $z \sim 2-3$ typically have $R_e$<1 kpc (e.g., van der Wel et al., 2014). Studies aiming to resolve these IMF radial gradients will unfortunately remain beyond reach, even with JWST's capabilities, due to its moderate spatial resolution (limited by the NIRSpec pixel size of 0.1 arcsec) and sensitivity (see Figs. 1 and 3). This unexplored regime is precisely where IMF measurements will offer the most powerful constraints on the origin and evolution of IMF variations across cosmic time.

#### 2.1.4. Ultra-Diffuse Galaxies

Ultra-Diffuse Galaxies (UDGs) are galaxies with sizes comparable to those of normal galaxies (e.g., the Milky Way), but approximately 100 times less massive (e.g., van Dokkum et al., 2015). Their origin remains an open issue and challenges current galaxy formation models, particularly concerning their dark matter content and formation histories. There are two main scenarios for their formations. The first assumes that they are failed massive galaxies whose star formation was prematurely quenched (e.g., $z$~2). Consequently, they formed dwarf-like stellar masses within relatively massive DM halos, thus appearing red, old, metal-poor, DM-dominated and featuring a rich globular cluster (GC) population, as indeed observed for many UDGs (e.g., van Dokkum et al., 2015; Toloba et al., 2023). The second main hypothesis suggests they were originally low-mass dwarf galaxies that have expanded in a low-mass DM halo due to feedback or environmental processes (e.g., Carleton et al., 2019). In this case, they should exhibit bluer colors, extended star formation histories, and host fewer GCs, like typical dwarfs, as observed for many UDGs (e.g., Toloba et al., 2023).

*Observational needs and challenges* - Probing the nature of UDGs requires a combined analysis of the kinematic and properties (age and metallicity) of GCs and field stars. This information would allow us to deduce the DM content of these systems, the star formation history, and the parameters of the stellar population, in particular metallicity, thus distinguishing between different formation channels. These types of observations require high angular resolution to disentangle crowded regions (see Fig. 3) and multiplexing to acquire spectroscopy for more than one target at a time, which is challenging both for NIRSpec due to the fixed slit width (0.2") of the MSA system, and for the ELT's AO-assisted spectrographs that provide single-object observations.

The issues related to UDGs are addressed in the articles by Mirabile et al. and by Iodice et al. in this special issue.

### 2.2. Population III stars

Population III stars (Pop III) are the first generation of stars predicted to form in the Universe. They arise from gas consisting exclusively of the primordial elements created during the Big Bang, namely hydrogen and helium, with no elements heavier than He (metals). No direct observations of PopIII stars have been made to date. However, their existence is supported by cosmological simulations (e.g., Abel et al., 2002; Maio et al., 2010; Park et al., 2021; Yajima et al., 2023), and the observation of extremely metal-poor halo stars, which are believed to be enriched by metals produced in PopIII stars (e.g., Hartwig et al., 2018; Vanni et al., 2023). The presence in a spectrum of prominent helium HeII($\lambda$1640) emission line and, eventually, Balmer and Ly$\alpha$ (if not absorbed) lines and no metal lines is believed a sign of extremely metal-poor conditions (e.g., Nakajima and Maiolino, 2022; Katz et al., 2023; Vanzella et al., 2023). Actually, the detection of high equivalent width HeII($\lambda$1640) emission would provide the definitive Pop III identifier. Trussler et al. (2023) show that for a $M^* \sim 10^6$ M⊙ PopIII galaxy at $z$~8, the detection of an EW~50 HeII($\lambda$1640) emission would require (ultra-)deep integrations (5–150 h) with NIRSpec/G140M, leaving little hope for the routine detection of Pop III populations with current instrumentation.

*Observational needs and challenges* - In fact, detecting Pop III also poses a challenge for the ELT, given the expected weakness of the underlying continuum and the emission line. In principle, given the extremely short life-cycles of these high-mass stars, the probability of observing them is maximized by searching in the earliest epochs of the Universe, e.g., within the first ~0.5 Gyr ($z$>8–9). Therefore, the detection of HeII($\lambda$1640) line at these redshifts requires the use of near-infrared spectroscopy ($\lambda$>1.6 μm). Assuming a mass of the order of $M^* \sim 10^6$ M⊙ PopIII stars, we can suppose that such stellar system has the size of GMCs, i.e., ~200 pc. Therefore, to maximize the probability of detecting HeII, the marker of PopIII, high sensitivity, near-infrared coverage, and finer angular resolution than NIRSpec to maximize the S/N are needed (see Figs. 1 and 3). A detailed derivation of the HeII[1640] line flux expected from a PopIII stellar system is given in Section 3.4.

### 2.3. Star formation and young stellar objects

The determination of the IMF in the very low-mass regime ($\ll$ 1M⊙) is a key unresolved issue in star and planet formation, particularly in low-metallicity (low-Z) environments (Zinnkann et al., 2024) and supermassive star clusters (M> $10^4$ M⊙) (Portegies Zwart et al., 2010).

Most of our knowledge on low-mass star formation comes from studying Young Stellar Objects (YSOs) in nearby star-forming (SF) regions ($d$ < 500 pc; e.g., Evans et al., 2009; Dunham et al., 2015), where spectroscopic surveys have addressed the relationship between accretion, jets, and disk structure (e.g., Alcalá et al., 2017; Nisini et al., 2018; Manara et al., 2021; Pittman et al., 2022; van Dishoeck et al., 2025). On the contrary, due to limited sensitivity and spatial resolution, little is known about YSOs in distant (kpc-scale) SF regions. In these distant regions, environmental factors like low-Z, very massive young clusters (M> $10^4\,M_\odot$) and intense radiation from massive stars may significantly affect accretion/ejection processes (e.g., Allen et al., 2025), and, ultimately, the final stellar mass (De Marchi et al., 2024).

A highly debated topic is whether low-Z environments accelerate the dispersal of circumstellar disks (Yasui, 2021) compared to solar-Z analogs. This rapid dispersal is theorized to be caused by higher mass accretion rates or more efficient photoevaporation (e.g., Winter and Haworth, 2022) driven by intense UV radiation from nearby massive stars. This would consequently shorten the time available for planet formation.

Conventional wisdom suggests that YSOs forming in low-Z and/or massive clusters should experience disk dispersal much earlier ($\leq$ 1 Myr), contrasting with the $6-8$ Myr timescales observed in low-mass, solar-Z clusters (Yasui et al., 2023). However, recent JWST studies hint at much longer disk lifetimes in low-Z SF regions (De Marchi et al., 2024). This conflict underscores the pivotal role the environment plays in circumstellar disk evolution and, ultimately, in the (debated) dependence of exoplanet frequency on stellar metallicity.

A detail case about YSO and the related issues above is described in the paper by Alcalá et al. in this special issue.

*Observational challenges* - Previous spectroscopic surveys of YSOs in distant environments, such as the outer Milky Way and the Magellanic Clouds (MCs), were largely limited to the brightest objects due to low sensitivity and difficulties with background subtraction. Even with the advanced sensitivity of JWST IR observations of very distant SFRs, interpreting the data remains challenging due to the potential for spectral contamination from unresolved multiple objects falling within the same observational aperture. Addressing these issues, crucial for accurately determining the IMF and understanding star formation in these extreme environments, needs high-angular resolution to disentangle crowded regions (see Fig. 3) and multiplexing capabilities for multiple objects.

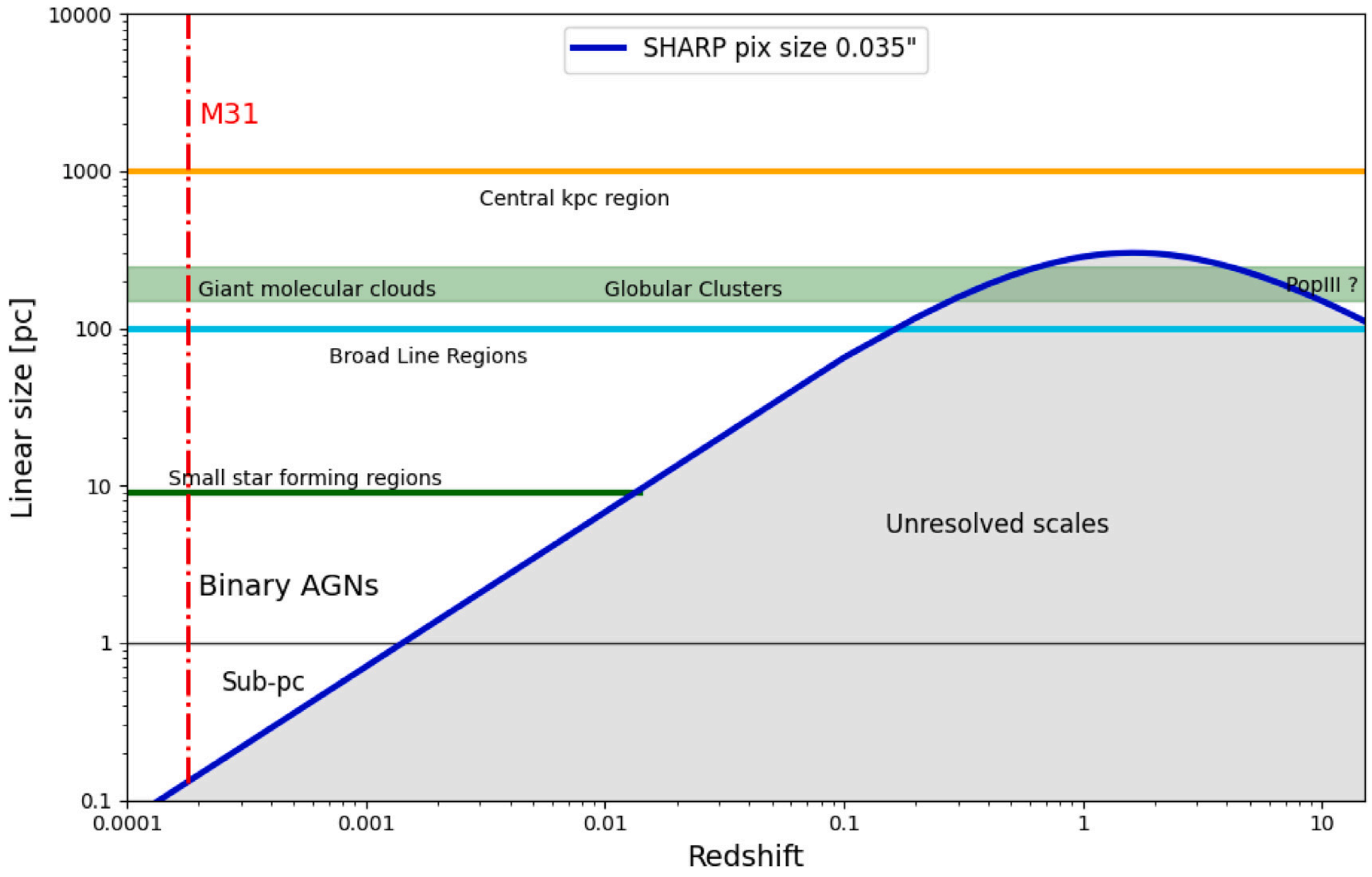

**Fig. 3.** Linear size [pc] in logarithmic scale subtended by a pixel of SHARP, $\theta$ = 0.035" (blue curve), as a function of redshift. The orange line marks 1 kpc scale, the green stripe represents the typical size of giant molecular gas clouds, 150–250 pc, globular clusters and high-redshift clumps observed by JWST. The green stripe also indicates the hypothetical PopIII system at $z \sim 10$ discussed in Sections 2.2 and 3.4. Cyan and green lines mark the typical sizes of Broad Line Regions (BLR) and small nearby star forming regions respectively. The vertical dot-dashed red line marks the spatial resolution that can be achieved at a distance similar to M31 galaxy.

## 3. Main requirements

To effectively tackle the above issues, observations require a specific combination of capabilities: near-IR coverage [0.95–2.45]μm, high-angular resolution, and multiplexing capabilities. It is precisely these requirements that make SHARP intrinsically versatile, enabling a range of other equally important studies.

### 3.1. Near-IR coverage

Understanding the physics driving galaxy evolution requires tracking galaxy properties across cosmic time. This is achieved by analyzing specific spectroscopic atomic emission and absorption lines encompassing gas and stellar kinematics, chemical abundances and stellar population age.

Among the main absorption features, magnesium Mgb[5120] and various iron Fe features (e.g., Fe[4046, 4383, 5003]) provide information on stellar total metallicity ($Z$). Crucially, the Mg/Fe ratio is a diagnostic for the duration of star formation (e.g., Matteucci and Tornambe, 1987). This is because Mg is rapidly released (∼ few Myr) primarily by core-collapse Supernovae (Type II, Ib, Ic), while Fe is released on a much longer timescale (> 1 Gyr) predominantly by Type Ia Supernovae. The amplitude of the continuum break at 4000 Å (D4000) is sensitive to the age of the bulk stellar population (e.g., Bruzual A., 1983), while CaH&K[3950] absorption lines are excellent tracers of stellar kinematics.

H$\alpha$ and OII[3727] are strong tracers of star formation activity (e.g., Kennicutt, 1998). Combined with OIII[5000] and H$\beta$, they provide a diagnostic tool for the presence of Active Galactic Nuclei (AGN).

Ly$\alpha$[1216] line traces neutral hydrogen. In emission, it provides galaxy redshift and a rough estimate of ongoing star formation. In absorption, it provides a guess of the neutral hydrogen fraction in the IGM, tracing the progress of Cosmic Reionization. HeII[1640] emission line is considered a robust spectral signature for the presence of ultra-hot Population III (Pop III) stars in the early Universe (see Section 2).

Most of these key features are redshifted into the near-IR regime when observed in distant galaxies. Specifically, wavelengths longer than $\lambda > 0.9$ μm and $\lambda > 1.8$ μm are needed for galaxies at $z > 0.8$ and $z > 2.5$, respectively (see Fig. 2) besides a spectral resolution R = $\lambda/\Delta\lambda$ >1000

Crucially, these wavelengths ($\lambda$>1.8 μm) are needed to search for Population III stars, whose extremely short lives limit their detection to the first 0.5 Gyr ($z$>9-10) of cosmic time. Finally, near-IR is also crucial to probe dust-shrouded regions such as nearby star-forming regions to study star and planet formation, the nuclear regions of galaxies to study the physics of AGN and to resolve stars in the dust-shrouded center of Globular Clusters (GCs) of nearby galaxies.

Since we are dealing with observations from the ground, $\lambda_{lim}\simeq$2.45 μm is the optimal limit in the near-infrared, given that sky transmission is still high and sky emissions can still be efficiently removed.

### 3.2. Angular resolution

To investigate into the baryons assembly in galaxies, regions comparable to the physical sizes of Giant Molecular Clouds (GMCs) should be resolved over the whole cosmic time. Indeed, GMCs containing even more than $10^6$ M⊙ of molecular gas, are

- the main star-forming units
- the main metal production units
- galaxy rotation and kinematics tracers
- the cradles of globular clusters (GC) and, possibly,
- cradles of the first PopIII systems.

Typically, GMCs have sizes of 150 to 250 pc in nearby galaxies. An angular resolution of approximately 30–35 mas is necessary to ensure that these physical sizes are sampled over the whole cosmic time. This is shown in Fig. 1 where the linear size subtended by an angle of ∼0.035 arcsec is shown as a function of redshift. This angular resolution ensures to resolve sub-galactic regions where massive star formation, enrichment and quenching processes take place, along the whole cosmic time.

Therefore, MCAO-supported observations coupled with a pixel scale of ∼ 30 mas/pixel allows instruments like SHARP to dissect the fundamental building blocks of galaxies, enabling direct comparisons between the observed properties of star-forming regions at high redshift and local Universe.

Fig. 3 is similar to Fig. 1 but in logarithmic scale to better visualize the scales that SHARP can resolve at the different redshifts. It is worth to note that, in the local Universe, for distances comparable to, e.g. M31, an angular resolution of ∼0.03 arcsec would allow sampling scales of the order of 0.1 pc (see Fig. 3), allowing us to study the physics of the star formation in the local dusty star-forming regions.

### *3.3. Multiplexing*

High-angular-resolution observations should be performed simultaneously for many targets (multiplexing), both for high-redshift studies and for studies of the local Universe. At high redshift, multiplexing is needed to study, e.g., clusters of galaxies, proto-clusters, overdensities, and the multiple clumps seen by JWST in the early Universe, to derive their dynamical state and reconstruct their assembly history. In the local Universe, multiplexing would allow us, for example, to address the problem of the formation of GCs by estimating the metallicity and measuring efficiently the kinematics of their stars. This, in turn, would also allow us to search for the presence of intermediate massive black holes (MBH) with masses in the range $10^2$ – $10^4$ M⊙ expected, if any, in GCs.

### *3.4. Sensitivity*

To define the sensitivity requirements for SHARP, we considered two challenging scientific measurements: the spectroscopic measurements of radial gradients of the stellar population properties in high-redshift galaxies and the detection of the HeII[1640] line of a hypothetical PopIII stellar system. Addressing these two cases will allow SHARP to unlock a wide range of other issues based on weak continuum features or line emission diagnostics.

*Stellar mass assembly: surface brightness limit* - Investigating the presence or absence of radial gradients in the properties of the stellar population of galaxies allows us not only to assess whether star formation, chemical enrichment, and quenching mechanisms as well as the IMF are driven by local environmental properties or global processes, but also to test for the 'two-phase' scenario of galaxy stellar mass assembly (e.g., Oser et al., 2010). Checking for radial gradients requires measuring stellar population properties both in the dense central core, i.e. within half of the effective radius ($R_{core}$ = $R_e$/2) and in the fainter peripheral regions, i.e., at R≥2$R_e$. Furthermore, carrying out these spatially resolved spectroscopic measurements down to stellar masses of $\log(M_*/M_\odot) \geq 10$ at $z > 2-3$ is crucial for bridging the gap with the local Universe. At these early epochs, galaxies experience physical conditions and environments drastically different from those observed today. Reaching such mass limits enables a direct and homogeneous comparison with local studies and, most importantly, allows us to sample the critical mass thresholds around $\log(M_*/M_\odot) \simeq 10.3$ and $\simeq 11.2$, where fundamental scaling relations, such as the mass-size and mass-metallicity relations show significant changes in their slopes (e.g. Gallazzi et al., 2005; Cappellari, 2013; Saracco et al., 2017).

A typical passive galaxy at $z$∼3 with mass log(M*/$M_\odot$)∼10 has $R_e$∼1–1.5 kpc (0.12"–0.18"), $H_{AB}$ ∼23.5 mag, H-K∼0.2–0.3 mag and is characterized by a Sersic profile with $n \sim 3$ (e.g., Shuntov et al., 2025; Huertas-Company et al., 2025; Gargiulo et al., 2026). Here we will consider $K_{AB}$∼23.5 and $R_e$ = 0.12" as reference values. Half light falls within $R_e$ and, for a Sersic profile with $n = 3$, ∼30% of the light (i.e. $K_{core}$∼24.8) falls within $R_{core}$ = R $_e$/2 = 0.06" (∼500 pc), corresponding to surface brightness $\mu_{K_{core}}$ ∼20.1 mag/arcsec$^2$. Note that these central regions would be sampled from 9–10 pixels with NEXUS and VESPER, allowing investigations on much smaller spatial scales (if the flux allows it), while with NIRSpec at JWST from only one pixel. At 2$R_e$ the surface brightness fades by about 1.6 mag, i.e. to $\mu_K$ ∼ 21.7 mag/arcsec$^2$. Therefore, a surface brightness of about 22 mag/arcsec$^2$ assures to carry out measurements of stellar population gradients in the population of passive galaxies with log(M*/$M_\odot$)> 10 at $z$∼3. A reliable measurement of the main absorption features requires a S/N∼10 on the continuum. This translates to a limiting surface brightness $\mu_{K_{lim}}$ ∼23.0–23.5 mag/arcsec$^2$ at 1$\sigma$. This limit is within the observational reach of SHARP. Based on the SHARP Exposure Time Calculator (ETC), a 10-hour integration achieves this sensitivity with VESPER by integrating the signal over an annular region at $R = 2R_e$ with a radial thickness of 0.05" to 0.1".

*Detecting popiii stars: limiting line flux* - To define the SHARP line flux limit requirement, the line emission predicted for a PopIII stellar system has been considered, the importance of which has been discussed in Section 2.2 and whose detection is challenging for current instrumentation. The observational marker of the elusive Pop III systems is supposed to be HeII ($\lambda$1640) emission line produced by the transition n = 3–> 2. These systems should form within massive clouds of pristine hydrogen. Assuming a $10^{6-7}$ $M_\odot$ gas cloud, we can suppose that its size is about 200 pc, similar to GMCs, or smaller and that the star formation efficiency is ∼10%, resulting in a population of $10^6$ $M_\odot$ PopIII stars. The predicted HeII[1640] emission can be written as

$$\begin{aligned} L(He[1640]) &= E_{1640} \times N\nu_{3->2} \\ &= E_{1640} \times P_{3->2} \times Q(He^+) \end{aligned} \quad (1)$$

where $E_{1640} = hc/\lambda = 1.2 \times 10^{-11}$ erg is the photon energy at $\lambda$ = 1640 Å and $N\nu_{3->2}$ the number of photons produced by the transition n = 3–> 2 given by the product of Q($He^+$), the number of UV ionizing photons with energy > 54.4 eV, the ionization energy of $He^+$, and $P_{3->2}$ the probability of transition n = 3–> 2.

Schaerer (2003), using stellar population synthesis models at virtually null metallicity, estimates that a PopIII stars with Salpeter IMF in the mass range 1–500 $M_\odot$ produces ∼$10^{46}$ ionizing photons/s/$M_\odot$ (see Tab.1 Schaerer, 2003). Therefore, the expected ionizing photons for our PopIII stars is Q(He+) = $10^{46}$×$10^6$∼$10^{52}$ photon/s. Assuming Case B recombination regime, all ionizing photons produce recombination, a fixed fraction of which ($N\nu_{3->2}$ = $P_{3->2}$×Q($He^+$)) pass through the specific transition n = 3–> 2 generating the He[1640] line. The exact calculation of this fraction is a rather complicated quantum mechanical calculation that also depends on the plasma temperature and is beyond the scope of this estimate. For our purpose, we will consider that 50% of the recombinations pass through n = 3–> 2, that is 50% of the UV ionizing photons produces a HeII[$\lambda$1640] photon. Therefore, the expected luminosity from such a PopIII system will be $L(He[1640]) = 1.2 \times 10^{-11} \times 0.5 \times 10^{52} \approx 6 \times 10^{40}$ erg/s and the expected line flux for such a PopIII system at $z \sim 10$ is therefore $F(He[1640]) = L(He[1640])/(4\pi d_L^2) \approx 4 \times 10^{-20}$ erg/s/cm$^2$, where $d_L$ is the luminosity distance. If the emission originates from a region of about 200 pc or less, it is seen by SHARP like a point source,[4] sampled by one pixel independently of the light profile of the source. This limiting line flux can be reached by NEXUS in 10 h of exposure.

However, we highlight that the expected line flux depends on the assumptions made. Should the intrinsic emission be significantly weaker, these primordial systems might escape direct detection even with SHARP. Nevertheless, a non-detection would place tight upper limits on the mass of the system, on the efficiency of the star formation that would constrain the current Pop III formation models. Furthermore, this sensitivity, combined with high angular resolution, will unlock other critical science cases, from mapping the faint Ly$\alpha$ emission of the Intergalactic Medium (IGM) to uncovering the ultra-faint dwarf galaxies that possibly drove the cosmic reionization.

[4] At z = 10 the angular scale is ∼4.255 kpc/arcsec, hence 150–200 pc subtend an angle of about 0.035"–0.040"

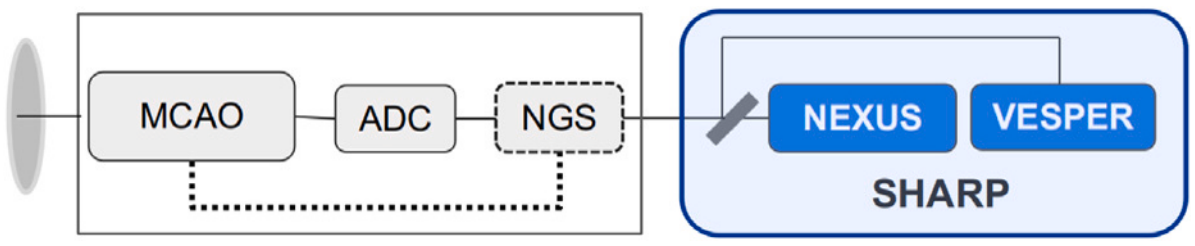

**Fig. 4.** Schematic view of the components and of the optical path from the last mirror of the telescope (gray mirror on the left) to SHARP. SHARP needs a MCAO unit, an NGS unit and an Atmospheric Dispersion Corrector (ADC). Note that the NGS unit may be integrated with the MCAO system.

## 4. SHARP

### 4.1. Instrument concept

Fig. 4 schematically shows the components from the telescope to SHARP. The SHARP spectrograph requires two fundamental systems to operate effectively: a MCAO system and an Atmospheric Dispersion Corrector (ADC). In the context of the ESO-ELT, the MCAO system is MORFEO. The AO-corrected light coming from MORFEO passes through the ADC unit. The ADC is crucial since it allows SHARP to simultaneously collect spectra over the entire wavelength range, regardless of zenith distance or rotation angle by eliminating atmospheric chromatic aberration. The Natural Guide Stars (NGS) unit is a system external to MORFEO. Its wavefront sensors monitor up to three natural stars and provide real-time feedback to MORFEO. This information is used in conjunction with six laser guide stars (LGS) to actively compensate for atmospheric turbulence a field of 160″ in diameter (see Fig. 5). The result of this correction is ideally an AO PSF uniform across the entire field. In reality, the spatial performance and uniformity of an MCAO system like MORFEO depend on atmospheric conditions (e.g., the turbulence profile) and, crucially, on the geometry and brightness of the NGS constellation. MORFEO delivers its optimal and most uniform performance when three bright (e.g., H<17) and well-spaced NGSs are available. The proximity of their position to the center of the field and their spatial distribution, ideally approximating a symmetric layout, are fundamental for an accurate tomographic reconstruction (see e.g., Arcidiacono et al., 2026, for a discussion). If the constellation is asymmetric, relies on fainter stars, or is limited to only two NGSs, the uniformity of the correction degrades. In extreme cases where only a single NGS is available, the system loses plate scale control (e.g., Plantet et al., 2022). This tends to blur the PSF over longer exposures and consequently reduces the concentrated energy within a given spectroscopic slit or single spaxel (e.g., Carlà et al., 2022). The end result is a spectrum with lower S/N and a partially loss of spatially resolved information.

The MORFEO corrected field encompasses the full field of view of SHARP's spectrographs. Notably, the NGS unit used by SHARP shares its design with the unit feeding the MICADO instrument, albeit without the SCAO (Single-Conjugate Adaptive Optics) module, which is not required for SHARP.

We note that for Integral Field Unit (IFU) observations, atmospheric dispersion correction can be effectively achieved via post-processing. Consequently, a removable ADC placed in the beam during VESPER observations could potentially enhance sensitivity and optimize the overall performance of SHARP. However, the conceptual study of the ADC is beyond the scope of this current paper.

We remind that, for ground based observations, an AO PSF is composed of two components: a diffraction-limited (dl) core, whose

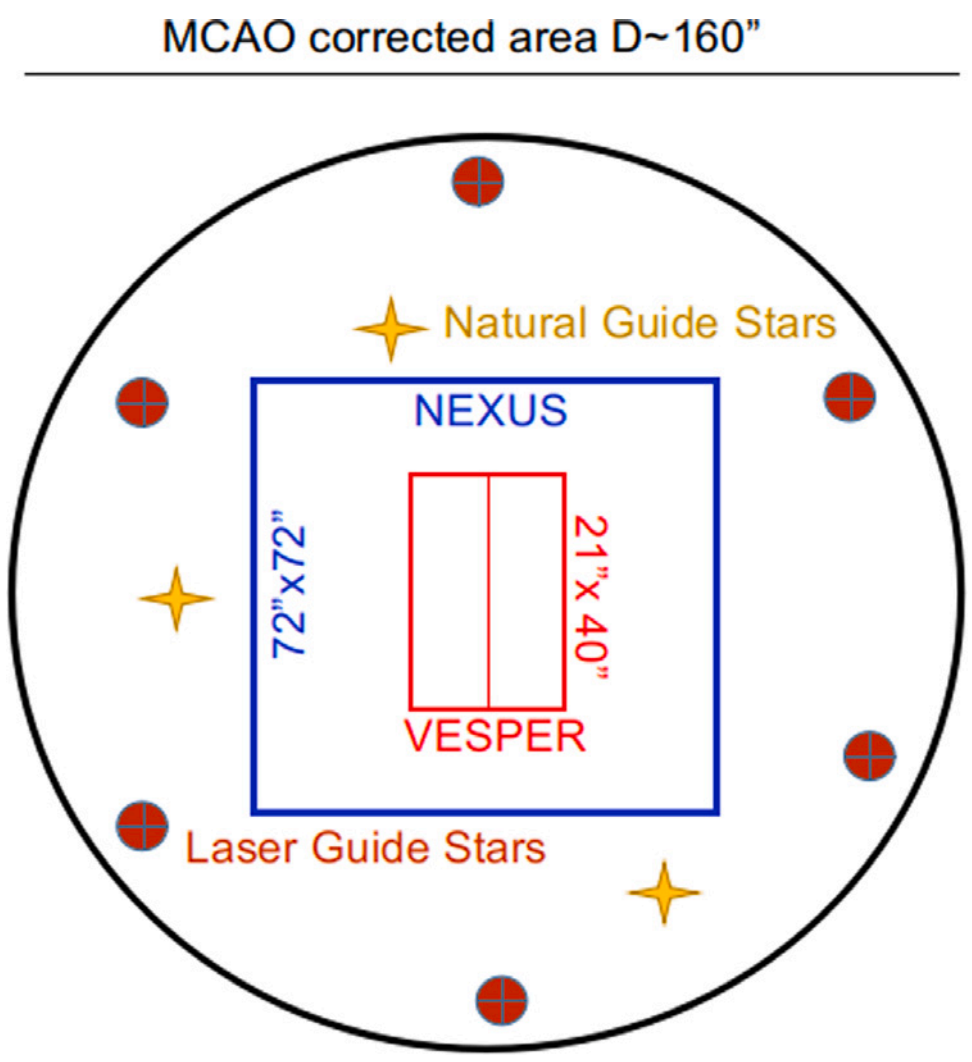

**Fig. 5.** The Field of View (FoV) of NEXUS (blue square, 1.2'×1.2') and the area probed by the 12 FSs of VESPER (see 4.4, red rectangle, ~21"×40") are shown on the AO-corrected area of the MCAO unit MORFEO (black circle, diameter D~160"). The dark-red filled circles are the wavefront sensors for the 6 Laser Guide Stars used by MORFEO, while the yellow stars are those for the 3 Natural Guide Stars.

FWHM depends on the aperture D of the telescope according to the relation $FWHM_{dl} = 1.22\ \lambda/D$, or

$$FWHM_{dl}(\lambda)[\text{mas}] = 206.3 \times \lambda[\mu\text{m}]/D[\text{m}] \tag{2}$$

and a seeing limited halo, whose extension depends on the seeing. For the ELT the diffraction-limited core has a $FWHM_{dl} \simeq 12$ mas at 2.2 μm. The Strehl ratio (SR), defined as the ratio of the intensity peak of the effective PSF to the intensity peak of the ideal diffraction-limited PSF, is a measure of the effectiveness of the AO system and represents, roughly, the flux fraction contained by the diffraction-limited component. For a point source and ideal AO system SR = 1. For comparison, for an MCAO system SR~ 0.5 under excellent seeing conditions and optimal NGS brightness and distribution on the field.

In the case of a space telescope like JWST, the three modules MCAO, NGS and ADC are not necessary, as they serve to correct for the effects of the atmosphere on the optical path of the light. The FWHM of the PSF is diffraction limited, dependent on the telescope aperture, and is uniform across the entire field of view of the telescope, barring distortions due to the optics. JWST has a $FWHM_{dl} \simeq 70$ mas at 2.2 μm.

### 4.2. SHARP architecture

Fig. 6 shows schematically the architecture of the instrument. SHARP consists of two main units: NEXUS, a slit-lets Multi-Object Spectrograph (MOS), and VESPER, a multi-Integral Field Unit (mIFU). Given the operative wavelength range SHARP is enclosed in a cryogenic tank. The light coming from the telescope and corrected for atmospheric dispersion by the ADC, is intercepted by the Unit Selector System (USS) placed onto the focal plane of the MCAO unit, just below the entrance window of SHARP. The USS hosts the Configurable Slit System (CSS) feeding NEXUS, and the Field Selector System (FSS) feeding VESPER. The USS switches the light between the two units by placing the CSS or the FSS under the entrance window.

SHARP hosts twelve cameras. Four cameras are used by NEXUS to simultaneously cover the whole wavelength range 0.95-2.45 μm (see Section 4.3); eight cameras feed VESPER that, in the current configuration, is composed of two modules, four cameras per module. The two modules support 12 probes, 6 per module (see Section 4.4).

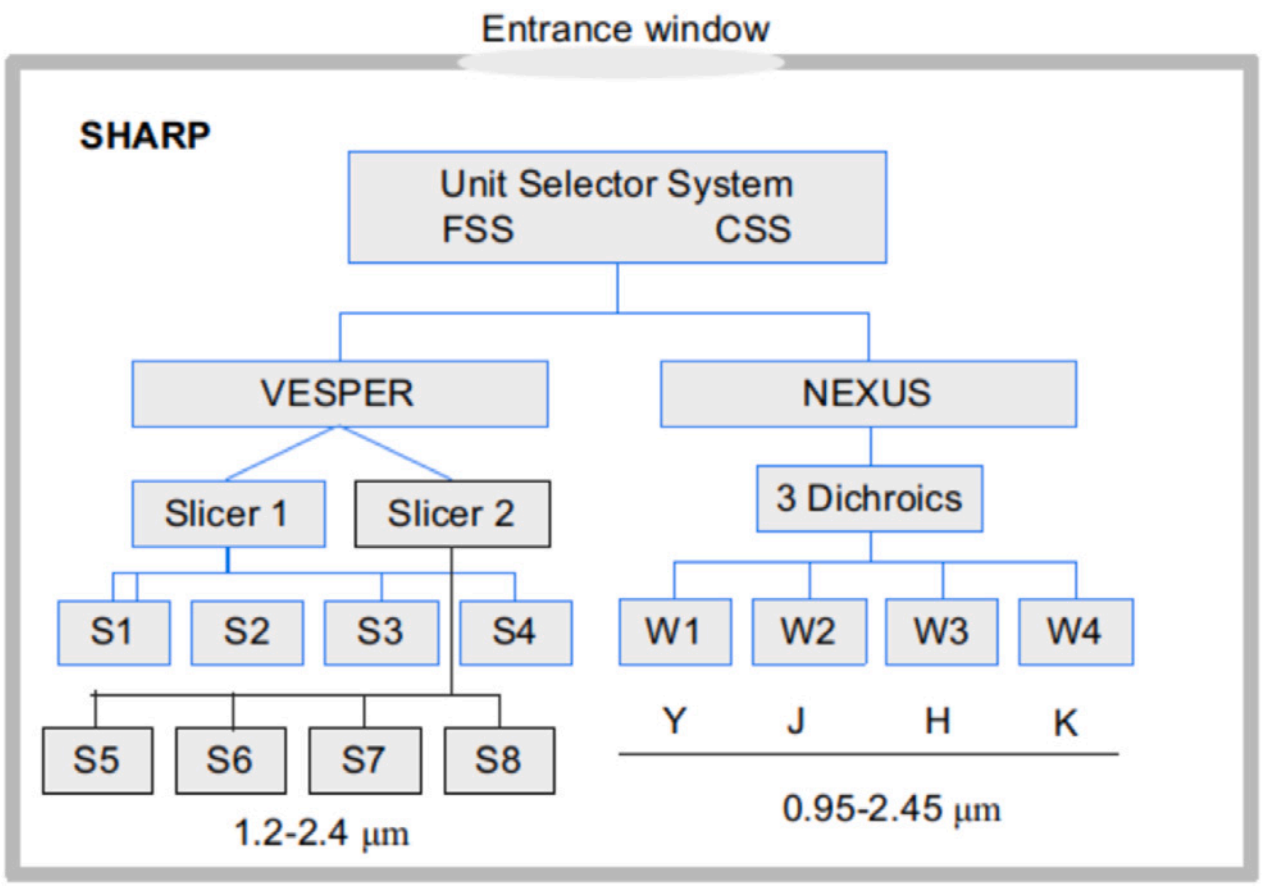

**Fig. 6.** SHARP architecture. SHARP is enclosed in a cryogenic tank. Below the entrance window of SHARP there is the USS that manages the light between NEXUS and VESPER. NEXUS - Three dichroics split the beam into four wavelength ranges feeding 4 cameras, W1-W4. The whole wavelength range 0.95-2.45 μm is thus simultaneously covered. VESPER - The image focused on the VESPER focal plane is divided into four equal stripes by the slicer each one fed by a camera (S1-S4). In the current configuration VESPER hosts two slicers to double the area probed.

In Fig. 7 it is shown the optical path from the entrance window of SHARP to the detectors. According to the USS position, the light beam can be intercepted by the CSS thus following the NEXUS optical path, or by the FSS hence following the VESPER optical path. A comprehensive description of the opto-mechanical design is given by Mahmoodzadeh et al. (2025).

### 4.3. NEXUS

*Overview and properties* - NEXUS, the Multi-Object Spectrograph (MOS), operates over an Adaptive Optics (AO) corrected field of view of approximately $1.2' \times 1.2'$ (35 mas/pixel; see Fig. 5). It is fed by a Configurable Slit System (CSS), which utilizes mechanical slit masks capable of deploying up to 30 slits of ~2.4" in length. The spectroscopic resolutions for extended sources, i.e., larger than the diffraction-limited PSF of ELT are R = $\lambda/\Delta\lambda$~6000, ~2000 and ~300 for a reference slit width of 0.2 arcsec. For point sources the resolution is R~17000, as defined by the FWHM of the diffraction-limited PSF (~12 mas at 2.2 μm)[5]

The different spectral resolutions reflect different needs. Accurate subtraction of OH lines generally requires a spectral resolution higher than R ~3000. However, not all scientific objectives demand high precision OH line subtraction. Some science cases benefit more from a higher Signal-to-Noise ratio (S/N), which is typically achieved with lower spectral resolution, rather than from meticulous OH line removal. The high resolution R~6000 is suited for studies requiring detailed spectral features, such as analyzing atomic line profiles (specifically their second and third-order moments) and making precise narrow line measurements. It provides the necessary accuracy for effective OH line subtraction. The intermediate resolution $R \sim 2000$ is preferred for studies focused on the mean properties of stellar populations and the physical properties of galaxies (e.g., stellar velocity dispersion, metallicity, and abundances). These investigations are less critically influenced by a less accurate OH lines subtraction but are strongly dependent on achieving a high S/N. The lowest resolution R~300 is thought to maximize the S/N of the continuum for extremely faint objects. Its primary purpose is to derive the overall spectral shape and detect continuum breaks and strong features in the faintest sources, where the fine subtraction of OH lines is not fundamental.

*Optical design and architecture* - Fig. 7 shows the optical path of NEXUS from the CSS to the four cameras together with the main optical elements (see also Mahmoodzadeh et al., 2025). A description of the CSS is presented later in the paper. The light exiting the CSS encounters a series of optics: three folding mirrors (FM) are used to provide a compact design. Three dichroic lenses (D1-D3) then split the beam into four distinct wavelength ranges: [0.95 – 1.15] μm, [1.15 – 1.45] μm [1.45 – 1.9] μm and [1.9 – 2.45] μm. Each range is directed to a dedicated camera (W1-W4) optimized for that operating band. The whole wavelength range (0.95–2.45 μm) is thus covered simultaneously. Each camera includes a grism wheel supporting three grisms[6] that achieve the resolutions mentioned above ($R \approx 300, 2000, 6000$) for a reference slit width of $0.2''$. Furthermore, a cold stop pupil is implemented to limit the thermal background, feeding the two cameras that operate at wavelengths $\lambda > 1.45$ μm. No aspheric surfaces are present in SHARP.

Each camera within NEXUS is equipped with four 2k × 2k detectors, resulting in a total of 8000 pixels available along the dispersion direction. This pixel count is needed to capture the entire spectrum for all observed sources, regardless of their position within the field or the chosen spectral resolution[7]

In Fig. 8 it is shown the Encircled Energy (EE) distribution as a function of radial distance at wavelength ~2.2 μm (fourth channel) for the central position. As input, an ideal point source modeled as a delta function and a point source modeled with a Gaussian with FWHM equal to the ELT diffraction limit were considered. The comparison shows that the overall image quality is dominated by the optical systems that precede SHARP, rather than the optics of NEXUS. Considering the diffraction limit of the ELT, >90% of the flux fall within one NEXUS pixel (15 μm, ~35 mas).

#### 4.3.1. The configurable slit system

Fig. 9 shows the concept design of the CSS used in NEXUS, conceptually similar to the designs by Henein et al. (2004) and Spanoudakis et al. (2008) implemented in the cryogenic spectrographs MOSFIRE at the Keck telescope (McLean et al., 2012) and EMIR at the GTC telescope (Garzón et al., 2022). A slit is formed at the position of a selected object by translating two opposite bars toward each other. The two bars effectively block the light originating from outside the slit. Both the slit position and the slit width are adjustable and controlled with micrometric precision (Mahmoodzadeh et al., 2025). This system offers a significant advantage over the Micro-Shutter Array (MSA) system feeding NIRSpec on the JWST, because it allows for both the perfect and simultaneous centering of all observed targets and the dynamic adjustment of the slit width.

[5] The slit width (sw) seen by a point source is given by the diffraction limit of the telescope, dl = 0.012 arcsec (12 mas) for ELT at 2.2 μm. Considering a resolving power R(0.2")≃6000 for a slit width of 0.2", the resolving power for a generic sw is R(sw) = (0.2"/sw)R(0.2"). Therefore, for the diffraction-limited slit width of 12 mas R(0.012) = (0.2"/0.012")≃ 6000 ≃100000. However, the pixel size of NEXUS is 0.035". Therefore, onto one pixel fall three resolution elements (0.035"/0.012"~2.9) and the actual resolving power becomes R~100000/2.9~34000. Considering two pixels to accurately sample the signal (Nyquist criterion) the expected resolving power for a point source is ~17000

[6] We anticipate employing high-performance near-IR grisms, comparable to the Lightsmith grisms developed for the Subaru telescope (Tanaka et al., 2020). A description of these optical elements will be provided in a forthcoming paper dedicated to the instrument's optical components and design.

[7] To illustrate the reason of this number of detectors, consider an observation at R~6000 at 2.1 μm. The resolution $\Delta\lambda = \lambda$/R≃3.5 Å. To satisfy the Nyquist condition, each pixel must sample 3.7/2≲1.75 Å/pix. Therefore, to cover the wavelength range 1.9-2.45 μm ≳3150 pixel are needed (~3660 for 1.5 Å/pix). If the observed source is located precisely at the center of the NEXUS field, two 2k detectors would theoretically be sufficient. However, to acquire a full spectrum even for sources close to the edges of the field, for example those shifted ±35″ from the center, we need to account for an additional ±1600(1830) pixels. This brings the total required number of pixels up to 6400(7320) to achieve the full spectral coverage. Along the spatial direction, given the pixel size of 0.035"/pix, 2048 pixels are sufficient to cover the 72" field of NEXUS. Therefore, four 2k detectors per camera assure the full spectral and spatial coverage of the NEXUS field of view.

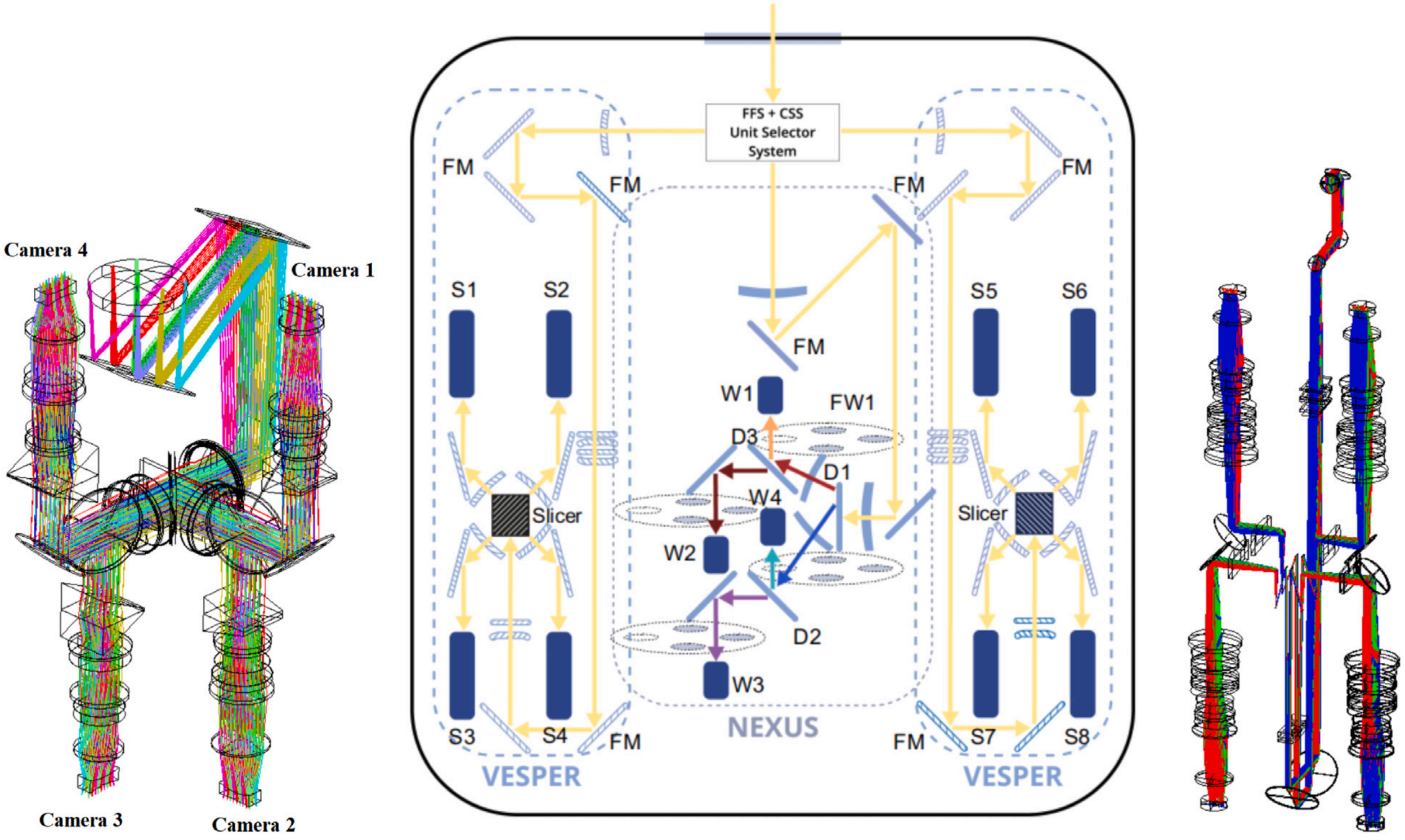

**Fig. 7.** Optical design from the focal plane of the MCAO unit to the detectors of NEXUS (left) and of one module of VESPER (right). In the center, the optical path from the SHARP entrance window to detectors. NEXUS - When the light is intercepted by the CSS, it passes through the slits to enters NEXUS. Three folding mirrors (FM) provide a compact design, while three dichroics (D1-D3) split the beam into four wavelength ranges each one fed by a dedicated camera (W1-W4). The whole wavelength range 0.95-2.45 μm is thus simultaneously covered. VESPER - When the light is intercepted by the FSS, it follows the path to VESPER. After some FM and collimators the beam is focused on the focal plane where it is placed the slicer. The image is divided into four stripes each one sliced at 0.031 arcsec and fed by a camera (S1-S4). A datailed description of the optical and mechanical components can be found in Mahmoodzadeh et al. (2025).

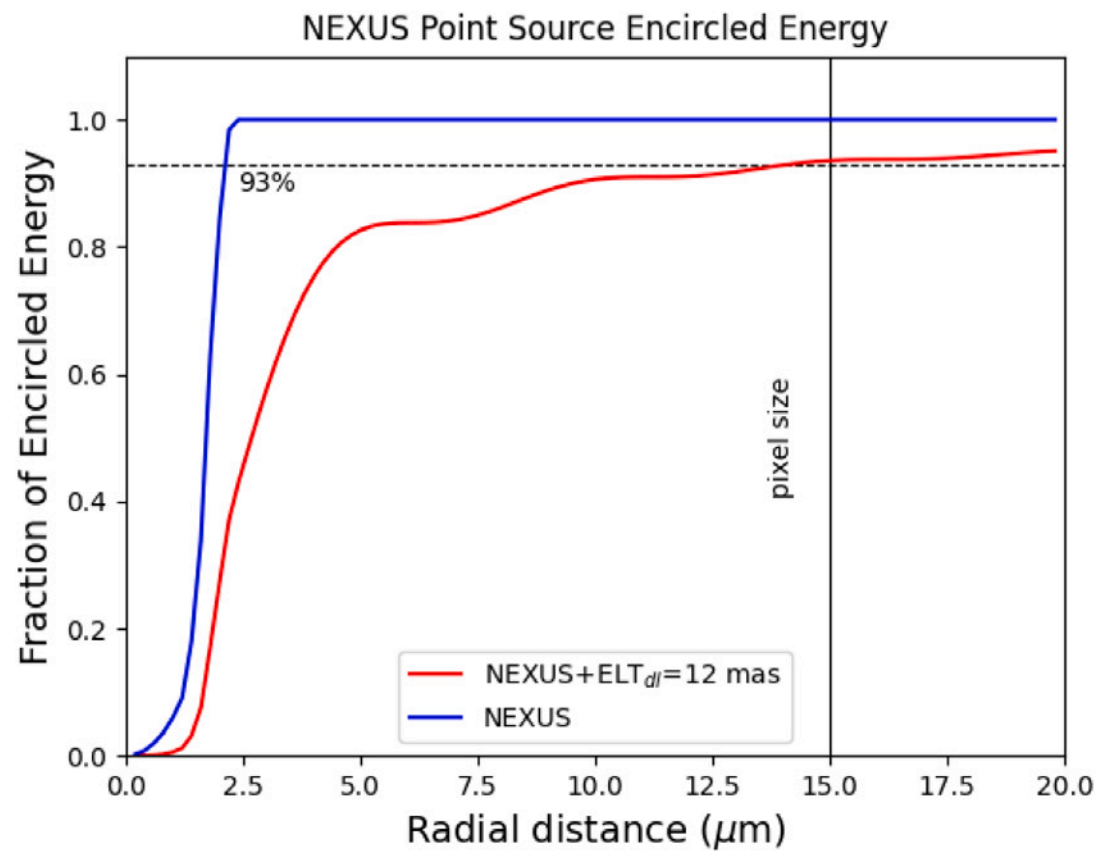

**Fig. 8.** Fraction of Encircled Energy as a function of radial distance for NEXUS at wavelength $\lambda = 2.19$ μm for a point source. The point source has been considered ideal (blue curve) and modeled accounting for the diffraction limit of the ELT (red curve). The pixel size of the detector is 15 μm.

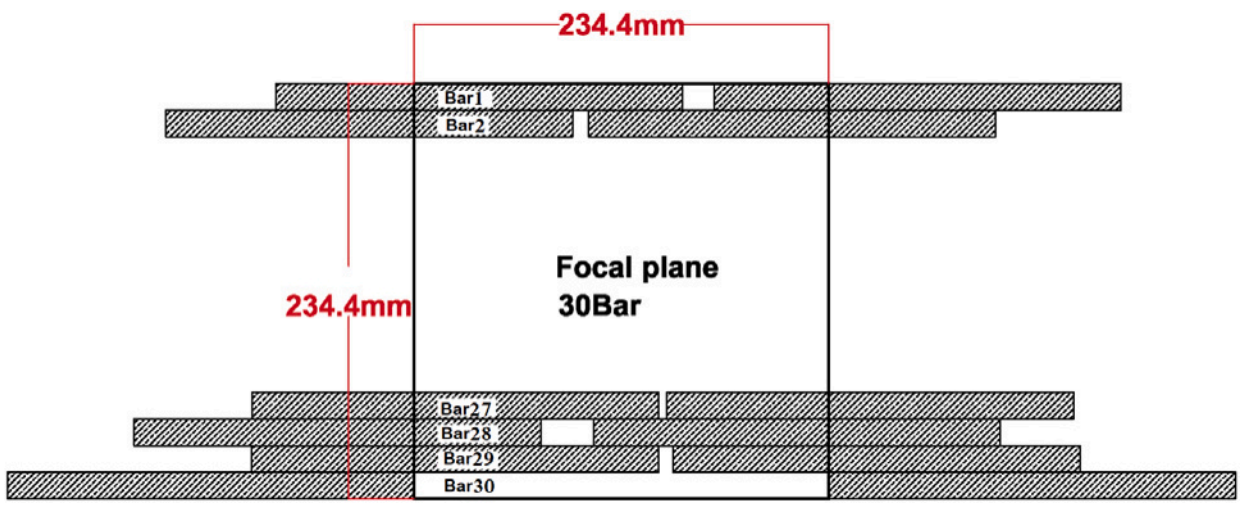

**Fig. 9.** Conceptual view of the configurable masking mechanism. Pairs of bars move toward each other to form a slit at the target position masking the outside regions.

In the current configuration, the minimum length of the configurable slit is ∼ 2.4″ (corresponding to a bar size of 7.5 mm), allowing dithering observations and a good sampling of the sky in case of observations of compact sources. This size limits the maximum number of simultaneously deployable slits to 30. Longer slits can be achieved by aligning two or more adjacent slits, resulting in a final slit length that is a multiple of 2.4″.

One of the goals of our concept design is to equip each NEXUS slit with an inversion prism (see Fig. 10, central panel). This would make NEXUS unique compared to any other slit system, configurable or not. The inversion prism (see e.g., Smith, 1990) can rotate the square field of ∼ 2.4″ × 2.4″ subtended by the slit by an arbitrary angle $\theta$, selected by the user, thereby aligning the target object precisely to the slit according to scientific needs. At the current conceptual design stage, an Abbe-König prism is considered as the baseline solution to control the orientation of the slit field (in Fig. 10 a Pechan prism is shown for rendering purposes only). We note that the physical impact of this prism in terms of throughput losses is negligible. Internal reflections inside the Abbe-König prism are virtually lossless, as they exploit the angle of incidence and occur at a constant refractive index. Consequently, the transmission penalty is dominated by the two air-glass interfaces (the entrance and exit faces). By applying high-performance broadband anti-reflective coatings capable of > 99% transmission per surface, the total reflective throughput loss introduced by the prism is kept below 2% (0.99 × 0.99). Given this minimal impact on throughput, we consider the prism to be a highly favorable trade-off, as it grants the unique operational capability fields to the slits according to the user needs.

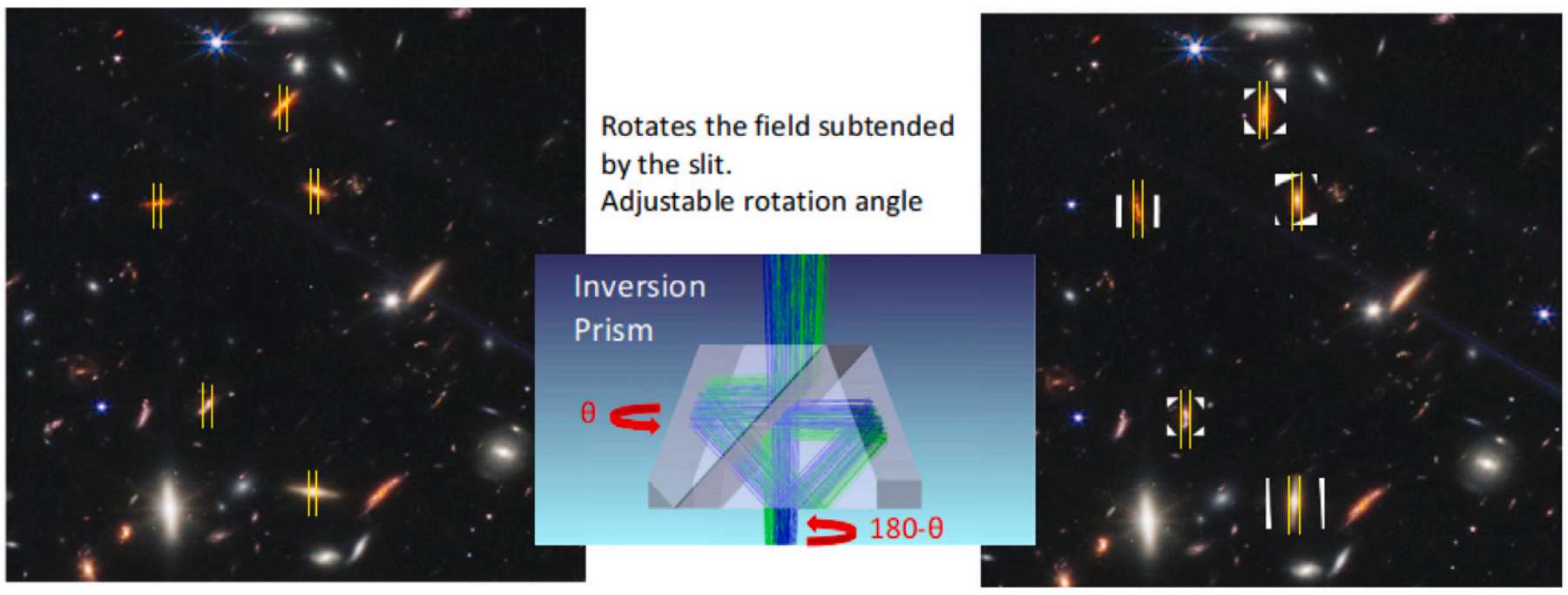

**Fig. 10.** Optical design of a Pechan inversion prism (central panel). This prism is composed of two paired prisms. The light entering from above undergoes five reflections and comes out from below inverted. Rotating the prism by an angle $\theta$ the coming out image is rotated by a similar angle. As example, all the target galaxies (left-hand panel) are aligned along the slit according to their major axis (right-hand panel).

Fig. 10 provides a compelling example of this application, illustrating how a sample of selected galaxies are all rotated to align their major axes with the slit. This capability provides, for instance, the simultaneous derivation of the rotation curve for every galaxy observed from which to derive their dynamical mass and their DM fraction. This feature makes NEXUS and hence SHARP truly unique.

### 4.4. VESPER

*Overview and properties* - VESPER, the multi-Integral Field Unit (multi-IFU) instrument, is an image slicer designed to simultaneously cover the wavelength range of 1.2–2.4 μm. The spectral resolution is set at R∼3000 for extended sources and R∼4000 for point sources. VESPER is built as a modular system. Each module contains six probes, called Field Selectors (FSs). In the current configuration, VESPER comprises two modules, totaling 12 FSs. Each individual FS has a field of view of approximately 1.7" ×1.5". The FSs are arranged in a Cartesian $xy$ grid, with virtually null separation among them along the $x$ direction. They can be deployed along ∼40” in $y$-direction to probe an AO-corrected area of approximately ∼21” × 40” (as shown in Fig. 5). The total equivalent area covered by the 12 FSs is $1.7'' \times 1.5'' \times 12 \approx 31$ arcsec$^2$. This area is mapped onto a total of eight 4k × 4k detectors, 4 detectors per module (see below). Fig. 12 shows one VESPER module, which is composed of six FSs. Note that the FSs can be arbitrarily arranged along the $y$-direction to form $n$ distinct fields, each with a FoV given by $(1.7'' \times 1.5'') \times k$, where $k$ represents the number of contiguous FSs positioned at the same $y$-coordinate. The spaxel scale is 31 mas, which maximizes also the S/N per spaxel at $\lambda = 2.2$ μm, according to the performance (encircled energy distribution) of MORFEO (see e.g., Arcidiacono et al. in this issue).

*Optical design and architecture* - Fig. 7 shows the optical path of VESPER from the FSS to the four cameras of each module (see also Mahmoodzadeh et al., 2025). A description of the FSS is presented later in the paper. The light exiting the FSS encounters a collimator and three folding mirrors (FM), a set of lenses to introduce a slight anamorphism and finally other two folding mirrors that send the light toward the slicer. The slicer, that will be described later in the paper, divides the image into four stripes each one fed by a camera (S1-S4). Unlike NEXUS, each camera covers the entire operating wavelength range of VESPER, i.e., 1.2-2.4 μm. Each camera includes a fixed grism (R∼3000) and it is fed by a 4k×4k detector. Also for VESPER, no aspheric surfaces are present.

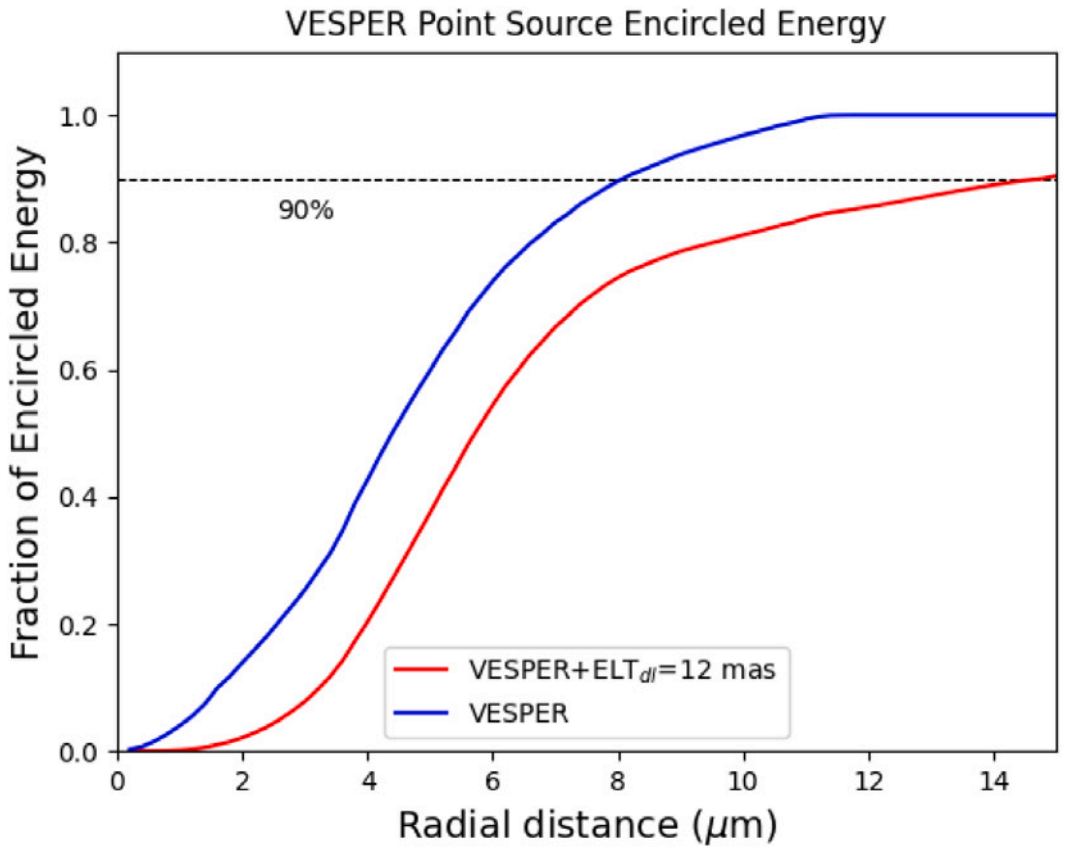

**Fig. 11.** Fraction of Encircled Energy as a function of radial distance for VESPER at wavelength $\lambda = 2.19$ μm. The two curves, as in Fig. 8, refer to a point source on one of the 288 mirrors of the slicer, in the ideal case of no diffraction (modeled as a delta function, blue curve), and modeled as a gaussian with a FWHM equal to the diffraction limit of the ELT (red curve).

Fig. 11 presents the encircled energy as a function of radial distance for a point source as seen by one of the 288 mirrors of the slicer (see below). The comparison is made between the ideal case without considering the diffraction limit and a more realistic case in which the point source is modeled as a gaussian with FWHM equal to the diffraction limit of the ELT. This comparison offers a direct measure of the quality of VESPER's optical design: the overall image quality is dominated by the optical systems that precede SHARP, rather than the internal slicer optics.

#### 4.4.1. The field selector system (FSS)

Fig. 12 shows the concept design of the Field Selector System (FSS) feeding one module of VESPER, hence covering half of the total area depicted in Fig. 5. The six FSs are aligned along the $x$-direction and can be moved by $\Delta y$ along the $y$-direction to the target position, sampling an area of approximately $10'' \times 40''$. Each FS is composed of a prism having the upper surface with power, and a collimator (Co) rigidly connected to each other (Fig. 12, center). Two 45-degree movable mirrors, M1 and M2, fold the beam. They move by $\Delta y/2$ each to maintain a constant optical path between the collimator Co and the camera Ca, specifically $\overline{CoM1} + \overline{M1M2} + \overline{M2Ca}$ = const. In this way, the optical path of all FSs remains identical, allowing them to share the same focal plane. On this shared focal plane, the six FSs form a combined rectangular image of size Y×X = 1.5" × [(1.7" × 6)]~1.5" × 10", as if they were contiguous (Fig. 12, right). Onto the VESPER focal plane there is the slicer.

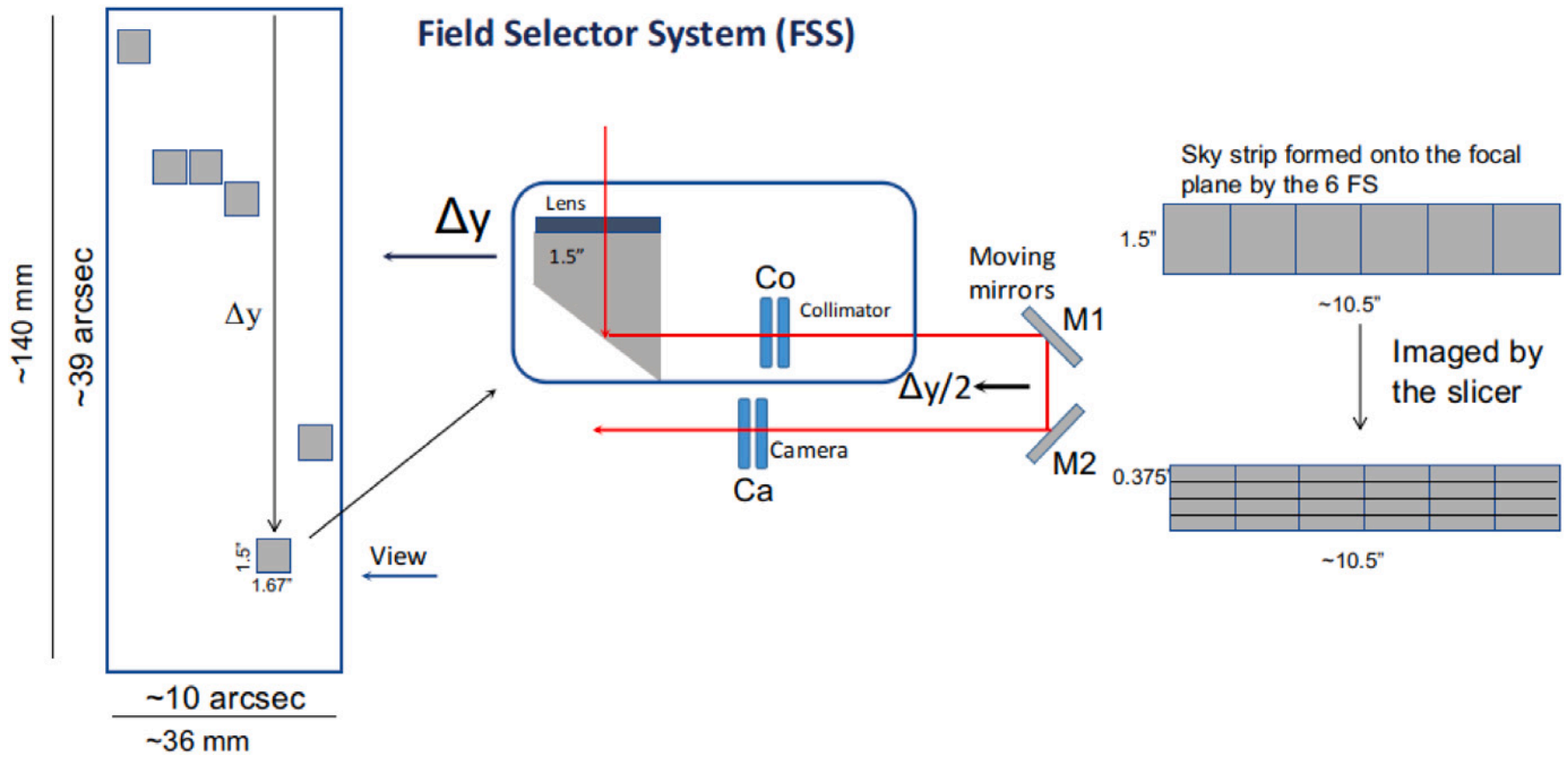

**Fig. 12.** Schematic view of the Integral Field Selector (FSS) system feeding one channel of VESPER. Left - The 6 FSs are aligned along the $x$-direction and can move along the $y$-direction sampling an area of ≈10"×40". Center - Each FS is composed of a prism and a collimator (Co). They rigidly move by $\Delta y$. Their movement is compensated by the movement ($\Delta y/2$) of the two 45-degree moving mirrors M1 and M2 to keep constant the optical path between Co and Ca. Right - The image formed onto the VESPER focal plane by the 6 FSs is a strip as if they were contiguous. Onto the VESPER focal plane there is the slicer.

The six FSs move along parallel planes alternatively offset up and down by about 25 mm. Furthermore, the prisms are oriented two by two to send the light in opposite directions. This design avoid interference and vignetting among the components (see Saracco et al., 2024; Mahmoodzadeh et al., 2025, for details).

### 4.4.2. The slicer

The light delivered by the FSS of VESPER is focused onto the slicer, which is conceptually similar to the slicer used in the MUSE spectrograph at the VLT (Henault et al., 2004). The slicer of VESPER is composed of four sets of 72 mirrors each (in comparison, the MUSE slicer uses sets of 48 mirrors). The mirror dimensions are approximately $1.0\ \mathrm{mm} \times 22\ \mathrm{mm}$. Fig. 13 illustrates the design of two such mirror sets. Each set of 72 mirrors samples a stripe equal to 1/4 of the image formed by the 6 FSs onto the focal plane along the wide direction, that is $0.375'' \times 10''$. The 72 mirrors are in turn divided into 6 groups of 12 mirrors, one group for each FS.[8] Therefore, 12 mirrors sample a stripe of $0.375'' \times 1.7''$, thus slicing it at 0.375"/12mirrors = 0.03125".

Corresponding to each set of slicer mirrors is a set of pupil mirrors (Fig. 13, lower panel) that converges the light toward the camera housing the spectrograph which generates a spectrum for every individual spaxel. The spectral resolution is R~3000 for extended sources and R~4000 for point sources. There is one camera dedicated to feeding each set of 72 mirrors, resulting in four cameras for a single VESPER module of 6 FSs. Consequently, the entire two-channel VESPER system utilizes a total of eight cameras and eight 4Kx4K detectors (one detector per camera). The expected throughput in K is 70%, neglecting the grism.

Table 1 summarizes the main properties of NEXUS and VESPER.

## 5. An illustrative science case for SHARP

In this section, we present an illustrative science case to qualitatively demonstrate the capabilities and versatility of SHARP. For a detailed quantitative treatment of specific science cases, we refer the reader to the individual articles comprising this special issue. The science case here considered represents an excellent case study because it encompasses multiple astrophysical features in a single target. This science case therefore serves a dual purpose: on the one hand, it clearly illustrates the complex observational requirements common to many astrophysical studies, and on the other, it demonstrates SHARP's unique capabilities compared to existing instruments to meet these requirements.

[8] Note that, for simplicity, in Fig. 13, only 4 groups of 12 mirrors are shown (48 in total) instead of the 6 groups that make up the set of 72 micro mirrors.

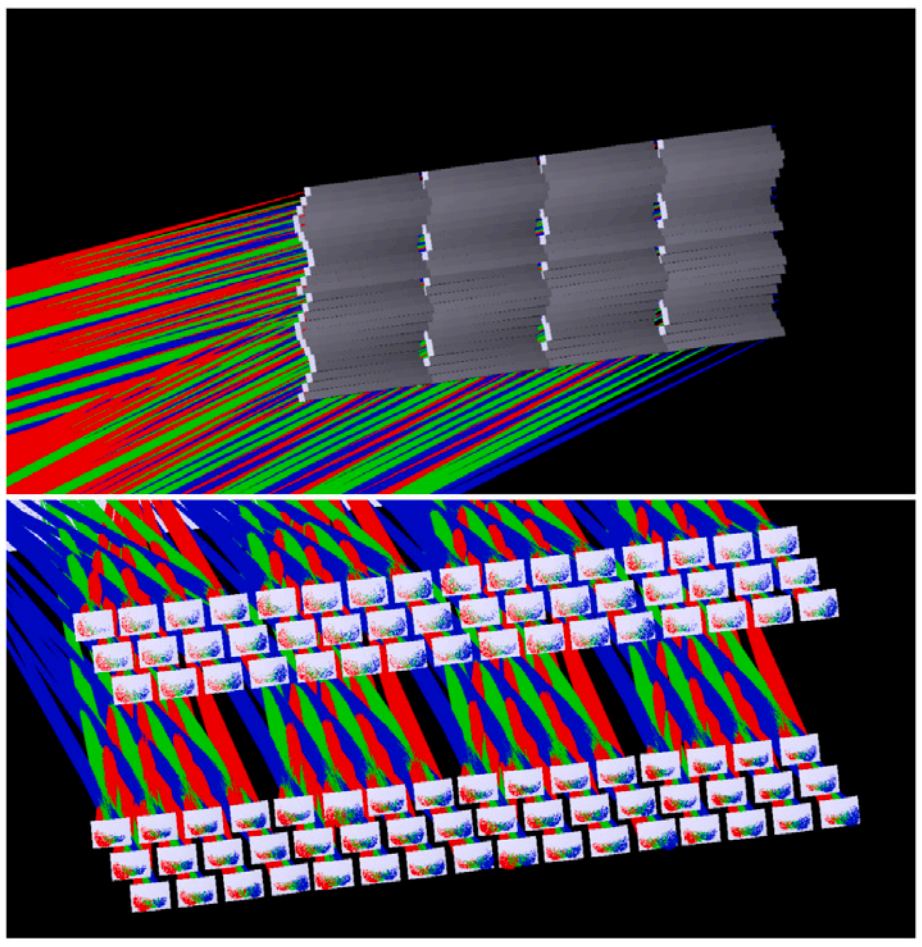

**Fig. 13.** Slicer onto the focal plane of VESPER. The slicer consists of four sets of 72 micro mirrors each and the corresponding sets of pupil mirrors. Two set of micro mirrors (upper panel) and of pupil mirrors (lower panel) are shown as example. Each set samples a stripe of 0.375"×10" of the image provided by the 6 FSs. The pupil mirrors bring the light to the corresponding cameras.

In Fig. 14 (left panel) it is shown the NIRCam image centered on galaxy GLASS-180009, as adapted from Fig. 5 in Bevacqua et al. (2026). This galaxy is at redshift $z$~2.66. The galaxy is quiescent, old (~1.7 Gyr) with respect to its redshift, with a stellar mass of about $4 \times 10^{10}$ M⊙ (Marchesini et al., 2023). It is probably member of an overdensity of galaxies whose extension is not yet defined. Bevacqua et al. (2026) show that most of stellar mass of GLASS-180009 formed within few hundreds Myr at $z$~11. Its effective radius is about 1 kpc. They detect a neutral gas inflow from the redshifted absorption of the NaI doublet at $\lambda\lambda$5890, 5896. They estimate a mass of gas $M_{gas}$~$10^8$ M⊙ and an inflow rate of about 19 M⊙/yr. Despite of the gas inflow, the galaxy appears quiescent with no signs of ongoing star formation (upper limit to the star formation of 0.2 M⊙/yr).

**Table 1**
Properties of the Multi-Object Spectrograph NEXUS and the Multi-Integral Field Unit VESPER of SHARP.

| SHARP property | NEXUS | VESPER |
|---|---|---|
| Spectral range (simultaneous) | 0.95-2.45 μm | 1.2-2.4 μm |
| FoV/Area probed | 1.2’×1.2’ | 21“×40” |
| Multiplexing | up to 30 slits (2.4“ slit length) | 12 FSs (1.7”×1.5" each) |
| Pixel scale | 35 mas | 31 mas |
| Spectral resolution (ext. source) | 300, 2000, 6000 | 3000 |
| Spectral resolution (point source) | ~17000 | ~4000 |
| Number of cameras | 4 | 8 |
| Number of Detectors | 4 per camera | 1 per camera |
| Matrix | 2k × 2k | 4k × 4k |

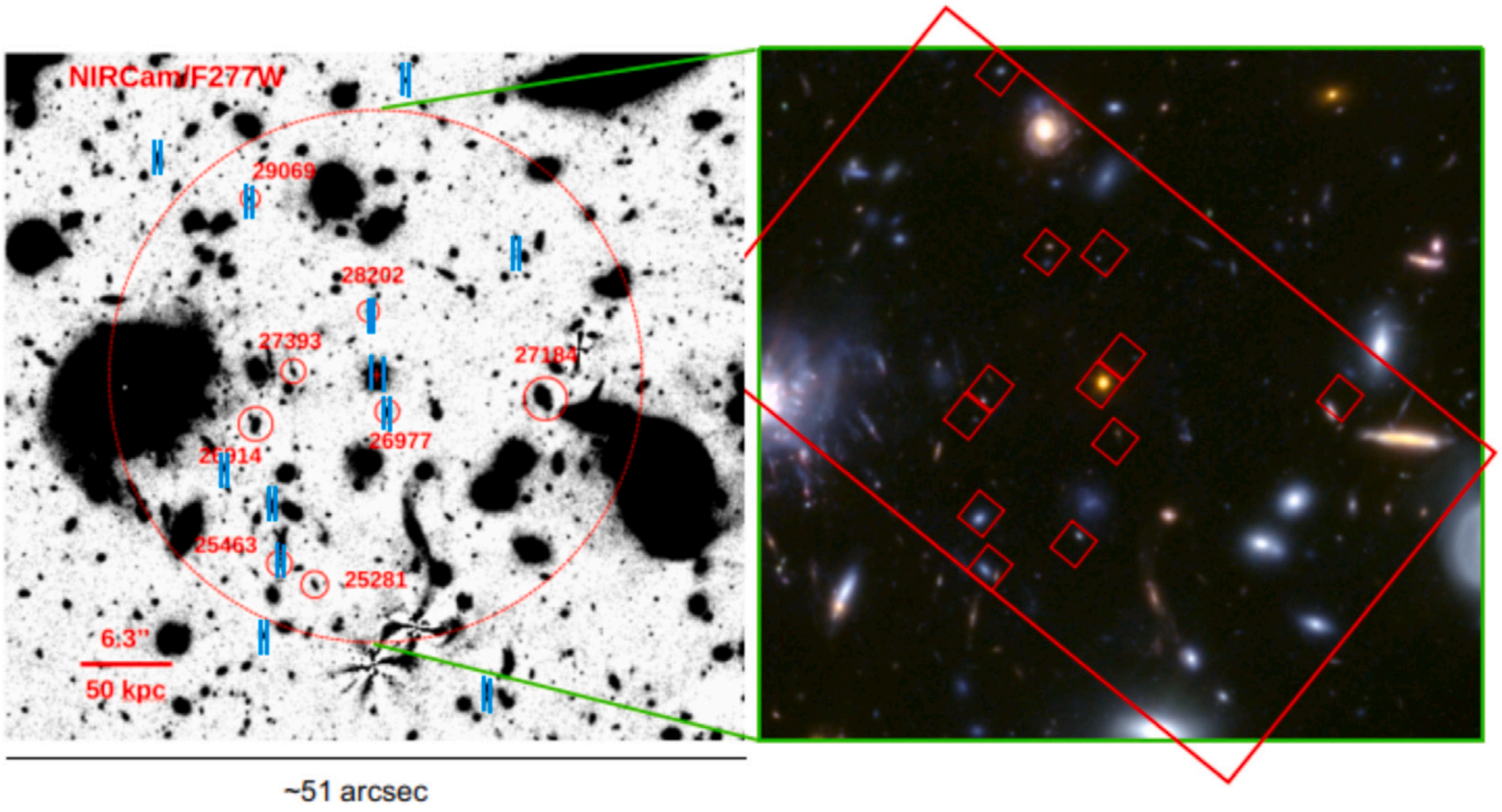

**Fig. 14.** Left - NIRCam image of the field centered on galaxy GLASS-180009 at $z$~2.66, adapted from Bevacqua et al. (2026) (Fig. 5). The field is about 51" × 51". The red small circles mark galaxies with similar redshift within a radius of about 150 kpc (large red circle). The FoV of NEXUS (72″ × 72″) fully encompasses the GLASS-180009 field. The light-blue small double-lines represent the slits of NEXUS, whose subtended field can be rotated thanks to the inversion prisms (see Section 4.3). Right - Composite JWST image (DAWN JWST Archive) of a square region of about 38″ × 38″ centered on GLASS-180009. The big red rectangle marks the area (∼ 20.5″ × 40″) probed by the 12 Field Selectors (FSs) of VESPER (represented by small red squares, each ∼ 1.7″ × 1.5″; see Section 4.4).

The nature of the gas inflow is unknown as is the reason why star formation is not occurring since at least the last Gyr (Bevacqua et al., 2026). The presence of the surrounding galaxies could suggest that the galaxy may accreting gas from nearby companions. On the other hand, the intergalactic medium (IGM) or cosmic filaments are also possible sources of incoming gas, since GLASS-180009 is likely be in an overdensity region.

To probe the nature of the inflowing gas, IFU observations of GLASS-180009 and the surrounding regions would be necessary. These observations would search for signs IGM or gas streams, as well as study the properties of the surrounding galaxies. MOS observations would be necessary to identify all galaxies at comparable redshift, defining their integrated properties, derive their kinematics, thereby defining the extension of the overdensity region (if any), characterizing the properties of its members, and ultimately constraining the properties of the overdensity itself.

An MCAO unit like MORFEO at the ELT uniformly corrects for atmospheric turbulence a field larger than that surrounding GLASS-180009 shown in the left panel of Fig. 14, where the slits of NEXUS (blue double-lines) targeting some surrounding galaxies are shown. Target galaxies can be aligned to the slits based on the necessary measurement (see Section 4.3). This represents a unique step forward in exploiting the high angular resolution over a wide field typical of MCAO systems, without which it would not be possible to obtain spatially resolved information for so many galaxies simultaneously.

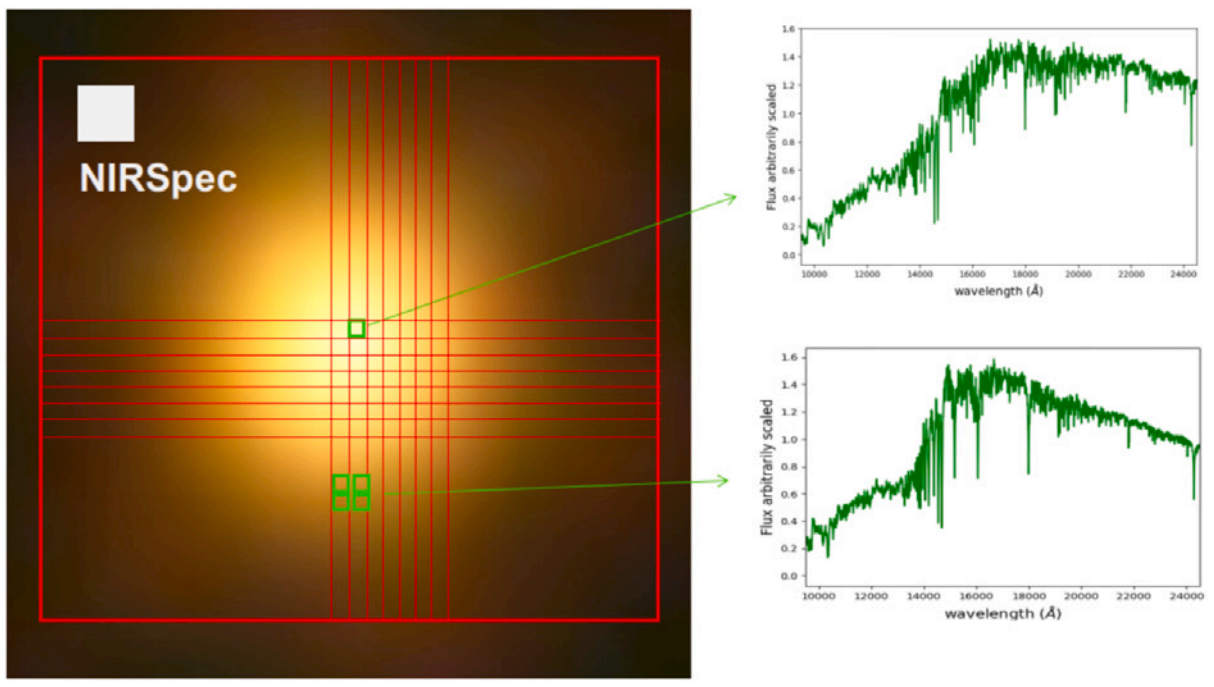

**Fig. 15.** Left - Zoom in of the composite JWST image centered on GLASS-180009. The big red square is the area (∼ 1.7″×1.5″) of a single FS of VESPER. The thin red lines schematically represent the slicing at 0.031". Highlighted in green are the central spaxel with a corresponding simulated spectrum representing a SSP 1.75 Gyr old and the sum of four spaxel in the outer region with a corresponding SSP of 0.9 Gyr old. The gray square represents the pixel size of NIRSpec (0.1").

In the right-hand panel of Fig. 14 it is shown a square region of about $38'' \times 38''$ centered on GLASS-180009. Superimposed is the area probed by the 12 FSs of VESPER ($\sim 20.5'' \times 40''$). The FSs are arranged to sample both the region close to the galaxy and some of the surrounding galaxies. This observation simultaneously probes the nature of the inflow, its relationship with the IGM (if present), and/or with surrounding galaxies.

Fig. 15 shows the field of the single FS (~1.7"×1.5") centered on GLASS-180009. The effective diameter of the galaxy (enclosing 50% of the light) is about 2 kpc (~0.25"), sampled by VESPER at about 250 pc (0.031"). The spatially resolved information, on these scales for the galaxy GLASS-180009, allows us to investigate the conditions of quiescence despite the available gas. Crucially, it enables the determination of gradients in the stellar population properties (age, metallicity, and possibly the IMF) and the kinematics, ultimately permitting the reconstruction of the galaxy's mass assembly history.

NIRSpec MSA and IFU can provide a sampling of 0.1" (~ 800 pc). The limited field of view (~3"×3") of the NIRSpec-IFU restricts it to single-galaxy investigation, thus preventing a full study of the inflowing gas and environmental connections. Different constraints affect the capabilities of HARMONI and MICADO. While HARMONI achieves an angular resolution comparable to that of SHARP, its limited field of view (up to $\sim 3'' \times 4''$) intrinsically restricts it to single-object studies. Similarly, the long-slit spectroscopic mode of MICADO is limited to single-object observations. Consequently, neither instrument can simultaneously probe the primary galaxy and its surrounding environmental connections. The crucial issue is that the full multiplexing potential of a large, uniformly corrected field at the ELT diffraction limit, as provided by an MCAO unit like MORFEO, is not exploited by the currently planned ELT spectrographs. Addressing this specific gap is the objective of SHARP.

SHARP would be highly complementary to the other upcoming ELT instruments. The fiber-fed MOSAIC (Pelló et al., 2024), for instance, is limited to 1.8 μm (preventing the detection of atomic features at $\lambda > 4800$ Å for galaxies at redshift similar or higher to GLASS-180009) and features a much lower angular resolution (fiber diameter $\sim 0.2''$) yielding integrated spectra. Nevertheless, MOSAIC is complementary because its fibers can be distributed over a much wider field of view than the MORFEO-corrected area, enabling large-scale surveys. Similarly, ANDES (limited to 1.8 μm; Marconi et al., 2024) will be complementary as a single-object spectrograph since it will offer an extremely high spectral resolution (R = 100000) that will be essential for detailed analysis of chemical abundances and kinematics of, e.g., IGM and exoplanets atmosphere.

## CRediT authorship contribution statement

**P. Saracco:** Writing – review & editing, Writing – original draft, Supervision, Resources, Project administration, Methodology, Investigation, Funding acquisition, Formal analysis, Data curation, Conceptualization. **P. Conconi:** Validation, Methodology, Formal analysis, Conceptualization. **H. Mahmoodzadeh:** Validation, Conceptualization. **C. Arcidiacono:** Supervision. **I. Di Antonio:** Supervision. **E. Portaluri:** Supervision. **P. Franzetti:** Supervision, Formal analysis. **A. Gargiulo:** Supervision. **I. Arosio:** Visualization. **L. Barbalini:** Supervision. **G. Lops:** Supervision. **E. Molinari:** Supervision. **J.M. Alcalá:** Supervision. **S. Bisogni:** Supervision. **R. Bonito:** Supervision. **E. Bortolas:** Supervision. **M. Cantiello:** Supervision. **A. Caratti o Garatti:** Supervision. **E. Cascone:** Supervision. **V. Cianniello:** Supervision. **E.M. Corsini:** Supervision. **F. Damiani:** Supervision. **F. D'Ammando:** Supervision. **E. Dalla Bontà:** Supervision. **M. Dall'Ora:** Supervision. **V. De Caprio:** Supervision. **G. De Lucia:** Supervision. **B. Di Francesco:** Supervision. **G. Di Rico:** Supervision. **V. D'Orazi:** Supervision. **C. Eredia:** Supervision. **A.R. Gallazzi:** Supervision. **M.G. Guarcello:** Supervision. **L. Izzo:** Supervision. **F. La Barbera:** Supervision. **M. Lippi:** Supervision. **M. Longhetti:** Supervision. **A. Longobardo:** Supervision. **C. Mancini:** Supervision. **M. Mirabile:** Supervision. **E. Piconcelli:** Supervision. **A. Pizzella:** Supervision. **L. Podio:** Supervision. **L. Prisinzano:** Supervision. **C. Tortora:** Supervision. **G. Vietri:** Supervision. **H.-F. Wang:** Supervision. **S. Zibetti:** Supervision.

## Declaration of competing interest

The authors declare that they have no known competing financial interests or personal relationships that could have appeared to influence the work reported in this paper.

## Acknowledgments

The SHARP team acknowledges support by Bando Ricerca Fondamentale INAF 2022, Techno-Grant "SHARP" - 1.05.12.02.01 and Bando Ricerca Fondamentale INAF 2024, Large-Grant "SHARP" - 1.05.24.01.01.

## Appendix. The SHARP science book

Beyond the illustrative example presented in Section 5 and the cases discussed in Section 2, a broader collection of detailed reference science cases is being compiled in the SHARP Science Book, accessible at https://www.sciencedirect.com/special-issue/10W3B6JJBXH. These cases encompass a variety of astrophysical themes, ranging from AGN and black hole growth and co-evolution (Severgnini et al., 2026; Vietri et al., 2026, Polletta et al. 2026), to galaxy evolution and environmental effects across cosmic time (e.g., Gargiulo et al., 2026; Mancini et al., 2026; Polletta et al., 2026), to very early Universe (Bisogni et al., 2026). The reader is referred to the dedicated papers within this Special Issue for a comprehensive analysis of individual programs. We note that the list above is incomplete since the compilation of this special issue is currently ongoing and some of the papers are still in the publication process.

## Data availability

No data was used for the research described in the article.